\documentclass[12pt]{article}
\usepackage{geometry}
\usepackage{titling}
\usepackage{authblk}
\usepackage{amssymb}
\usepackage{amsmath}
\usepackage{amsthm}
\allowdisplaybreaks[4]
\usepackage{slashed}
\usepackage{graphicx,color,epsfig}
\usepackage{multirow}
\usepackage{cite}
\usepackage{booktabs}
\usepackage[colorlinks=true,linkcolor=red,citecolor=blue]{hyperref}
\begin{document}
\title{Investigation of Effects of New Physics in $c\to (s,d)\ell^+\nu_\ell$ Transitions}
\author[1]{Xue Leng}
\author[2]{Xiao-Long Mu}
\author[1]{Zhi-Tian Zou}
\author[1,3]{Ying Li$\footnote{liying@ytu.edu.cn}$}

\affil[1]{\it \small Department of Physics, Yantai University, Yantai 264005, China}
\affil[2]{\it \small Institute of Particle Physics and Key Laboratory of Quark and Lepton Physics (MOE), Central
China Normal University, Wuhan, Hubei 430079, China}
\affil[3]{\it \small Center for High Energy Physics, Peking University, Beijing 100871, China}

\maketitle
\vspace{0.2cm}

\begin{abstract}
Recent anomalies in decays induced by $b\to c \ell^- \bar\nu_\ell$ transitions raise the question about such phenomena in the $D$ decays induced by $c\to (s,d)\ell^+\nu_\ell$ transitions. In the experimental side, current measurements on the pure leptonic and semileptonic $D$ decays agree with the standard model predictions, such agreements can be used to constrain the new physics (NP) contributions. In this work, we extend the standard model by assuming general effective Hamiltonians describing the $c\to (s,d)\ell^+\nu_\ell$ transitions including the full set of the four-fermion operators. Within the latest experimental data, we perform a minimum $\chi^2$ fit of the Wilson coefficient corresponding to each operator. The results show that the Wilson coefficients of scalar operators in muon sector are at the order of ${\cal O}(10^{-2})$, and others are at the order of ${\cal O}(10^{-3})$. The lepton flavor universality could be violated by the scalar operators. We also calculate the branching fractions, the forward-backward asymmetries and polarizations of final vector mesons and leptons with the fitted Wilson coefficients of scalar and tensor operators. It is found that the pure leptonic decays are very sensitive to the scalar operators. The effects of NP on the semileptonic decays with electron are negligible, while for the semileptonic decays with muon the effects of scalar operators will show up in the forward-backward asymmetries and polarizations of muon of $D \to P\mu^+ \nu_\mu$. The future measurements in BESIII and Belle II experiments will help us to test effects of NP and to further test new physics models.

\end{abstract}

\newpage
%%%%%%%%%%%%%%%%%%%%%%%
\section{Introduction}
%%%%%%%%%%%%%%%%%%%%%%%
Despite the discovery of the Higgs boson, the standard model (SM) is still regarded as the low-energy effective theory of a more fundamental one, because SM cannot explain the matter-antimatter asymmetry in the universe, it has not a dark matter candidate and it does not explain its own gauge group structure. So, one of the most important tasks in the particle physics community is searching for the new physics (NP) beyond SM, which can be examined via probing for NP signals directly at the higher energy colliders or testing SM with high precision at high intensity machines indirectly. As for the indirect approaches, the rare processes induced by flavor-changing neutral-current (FCNC) are generally considered to be an ideal plate to search NP, because FCNC only occurs by loops in SM and the branching fractions can be enhanced by new particles. However, in the past few years, there are some unexpected anomalies in the semi-leptonic $B$ decays induced by the charged current $b\to c\ell\bar\nu_\ell$, with respect to the corresponding SM predictions at the $(2-3)\sigma$ level (see e.g.~\cite{Li:2018lxi, Bifani:2018zmi, Watanabe:2017mip, Tran:2018kuv, Bhattacharya:2018kig}). It is very strange for us to find such large deviations from SM in these processes that occur at tree level. It should be emphasized that some anomalies might imply that the lepton flavour universality (LFU) is violated, which is the hint of the existence of NP, because the LFU is one of the major characters of SM. If these measurements can be further confirmed in the future experiments, which would be the signals of NP. Therefore, many NP models involving new particles have been proposed for explaining such tensions, such as $W^\prime$ models \cite{He:2017bft,Asadi:2018wea}, leptoquark models \cite{Celis:2012dk, Li:2016vvp, Celis:2016azn}, and models with charged Higgs \cite{Tanaka:1994ay,Iguro:2017ysu,Martinez:2018ynq}.

Since these anomalies were found in $B$ decays, it is natural to ask two questions. The first one is whether the similar discrepancy can show up in charmed mesons decays induced by $c\to (s,d)\ell^+\nu_\ell$, and the second one is whether the new introduced particles in the proposed models can affect the observables of these $D$ mesons decays \cite{Richman:1995wm, Ablikim:2019hff, Fajfer:2015ixa}. However, there is no any anomaly that has been observed in the experimental side so far \cite{Zhang:2019tcs}. In the theoretical side, a lot of efforts have been made in order to search for contributions of NP in $D$ decays in the past few years \cite{Dobrescu:2008er, Barranco:2013tba, Barranco:2014bva, Ricciardi:2016pmh}. We all know that the most important theoretical inputs are the heavy-to-light form factors, which are nonpertubative and can only be calculated by some nonpertubative approaches, such as approaches based on quark models, QCD sum rules and lattice QCD (LQCD). In particular, the results of LQCD is viewed as more reliable, as it is based on the first principles. Recently, the SM predictions of the semileptonic decays based on the LQCD \cite{Riggio:2017zwh} are in agreement with the world average experimental measurements \cite{Zyla:2020zbs} with large uncertainties from the CKM matrix elements. Such consistencies can be used to constrain the parameter spaces of NP \cite{Wang:2014uiz} or testing NP models. In the experiments, except the absolute branching fractions, there are many other observables such as the differential width, the forward-backward asymmetry and polarizations of final states, and most of them have not been measured now. So, in this work, we hope to fit the parameters of NP within the existed experimental data under single operator assumption and further check whether these fitted parameters can contribute to above observables. The comparisons between our results and future experimental data are helpful for probing the signals of NP.

In order to achieve the above purpose, we shall analyze all $D$ mesons decays induced by $c\to (s,d)\ell^+\nu_\ell$ in the model-independence manner, including the leptonic and semi-leptonic decays. Based on the general framework of the four-fermion effective field theory, we will perform a minimum $\chi^2$ fit of the Wilson coefficient of each operator to the latest experimental data. With these obtained Wilson coefficients, we present the predictions of other observables for testing in the future experiments.  For the pure leptonic $D$ decays, we will also study the LFU with current experimental data.

This work will be organized as follows. In Sec.~\ref{sec:framework} the framework in this work is presented, including the effective hamiltonian, the form factors and the helicity amplitudes. We list the parameters in Sec.~\ref{sec:paramerter}. The numerical analysis of leptonic decays will be investigated in Sec.~\ref{sec:leptonicdecays}, and the results and discussions of semi-leptonic decays of $D$ mesons are given in Sec.~\ref{sec:semileptonicdecays}, and the predictions of the physical observables are also given in this section, Finally, we give the summary and conclusions in Sec.~\ref{sec:summary}
\section{Framework}\label{sec:framework}
\subsection{Effective Lagrangian}
In particular, the energy scale of NP is supposed to be higher than the electroweak scale, thus the operator product expansion formalism (OPE) is reliable since it allows the separation between long-distance and short-distance interactions. In the OPE, the heavier degrees of freedom can be integrated out, resulting an effective Lagrangian where all high energy physics effects are parameterized by Wilson coefficients, namely the effective couplings multiplying the operators of the Lagrangian. In this spirit, considering all possible Lorentz structures, and assuming neutrinos to be left-handed, we then write down the effective Lagrangian for the decay $c \to q \ell^+\nu_l$ (with $q = d, s$) \cite{Huang:2018nnq, Mu:2019bin}
\begin{equation}
{\cal L}_{eff}=-\frac{4G_{F}}{\sqrt{2}}V_{cq}\Big[(1+C^{\ell}_{VL})O^{\ell}_{VL}+C^{\ell}_{VR}O^{\ell}_{VR}+C^{\ell}_{SL}O^{\ell}_{SL}
+C^{\ell}_{SR}O^{\ell}_{SR}+C^{\ell}_{T}O^{\ell}_{T}\Big]+h.c,
\end{equation}
where $G_{F}$ is the Fermi constant and $V_{cq}$ is the CKM matrix element. The four-fermion operators can be defined as
\begin{align}
O^{\ell}_{VL}=(\bar q\gamma^{\mu}P_{L}c)(\bar \nu_{\ell}\gamma_{\mu}P_{L}\ell),&\,\,
O^{\ell}_{VR}=(\bar q\gamma^{\mu}P_{R}c)(\bar \nu_{\ell}\gamma_{\mu}P_{L}\ell),\nonumber\\
O^{\ell}_{SL}=(\bar qP_{L}c)(\bar \nu_{\ell} P_{R}\ell),&\,\,
O^{\ell}_{SR}=(\bar qP_{R}c)(\bar \nu_{\ell} P_{R}\ell),\nonumber\\
O^{\ell}_{T}=(\bar q\sigma^{\mu\nu}P&_{L}c)(\bar \nu_{\ell}\sigma_{\mu\nu}P_{R}\ell),
\end{align}
where $P_{L(R)}=(1\mp\gamma^{5})/2$ and $C^{\ell}_{i}$ ($i = VL,VR,SL,SR,$ and $T$ ) are the corresponding Wilson coefficients at the scale $\mu = m_c$, with $C_{i} ^{\ell}= 0$ in SM. It should be noted that we assume all Wilson coefficients to be real for simplicity, i.e. that the NP effects do not involve new sources of $CP$ violation.
\subsection{The Form Factors}
In our calculations, the hadronic transition is parameterized by the heavy-to-light form factors, which are nonperturbative but universal. For the hadronic matrix elements of $D\to P$ transitions, $P$ denoting the pseudoscalar meson, they can be defined as
\begin{align}
&\langle P(p_{2})|\bar q\gamma^{\mu}c|D(p_{1})\rangle= f_{+}(q^2)\Big[(p_{1}+p_{2})^{\mu}-\frac{m_{D}^{2}-m_{P}^{2}}{q^{2}}q^{\mu}\Big]+f_{0}(q^2)\frac{m_{D}^{2}-m_{P}^{2}}{q^{2}}q^{\mu},\\
&\langle P(p_{2})|\bar qc|D(p_{1})\rangle= \frac{q_{\mu}}{m_{c}-m_{q}}\langle P(p_{2})|\bar q\gamma^{\mu}c|D(p_{1})\rangle=\frac{m_{D}^{2}-m_{P}^{2}}{m_{c}-m_{q}}f_{0}(q^2),\\
&\langle P(p_{2})|\bar q \sigma^{\mu\nu} c|D(p_{1})\rangle= -i(p^{\mu}_{1}p^{\nu}_{2}-p^{\nu}_{1}p^{\mu}_{2})\frac{2f_{T}(q^2)}{m_{D}+m_{P}},
\end{align}
where  $q^{\mu}=(p_{1}-p_{2})^{\mu}$ and two QCD form factors $f_{+}(q^2)$ and $f_{0}(q^2)$ encode the strong-interaction dynamics with satisfying  $f_{+}(0)=f_{0}(0)$. $m_{q}$ are the running quark masses.  For the various form factors of the $D\to K$, we adopt the latest results from the LQCD calculations \cite{Lubicz:2017syv, Lubicz:2018rfs}, and each form factor can be written as
\begin{align}
f_{+}^{D\to K}(q^{2})=&\frac{f^{D\to K}(0)+c_{+}^{D\to K}(z-z_{0})(1+\frac{z+z_{0}}{2})}{1-q^{2}/M_{D_{s}^{*}}}^2,\\
f_{0}^{D\to K}(q^{2})=&f^{D\to K}(0)+c_{0}^{D\to K}(z-z_{0})(1+\frac{z+z_{0}}{2}),\\
f_{T}^{D\to K}(q^{2})=&\frac{f_{T}^{D\to K}(0)+c_{T}^{D\to K}(z-z_{0})(1+\frac{z+z_{0}}{2})}{1-P_{T}^{D\to K}q^{2}},
\end{align}
where $z$ is defined as
\begin{align}
z=\frac{\sqrt{t_{+}-q^{2}}-\sqrt{t_{+}-t_{0}}}{\sqrt{t_{+}-q^{2}}+\sqrt{t_{+}-t_{0}}},
\end{align}
with $t_{+} $ and $t_0$ given by
\begin{align}
t_{+}=(m_{D}+m_{K})^{2},\,\,\,t_{0}=(m_{D}+m_{K})(\sqrt{m_{D}}-\sqrt{m_{K}})^{2},
\end{align}
and $z_0 =z(q^2=0)$. For the $D\to \pi$ transition, the scalar, vector and tensor form factors are also parameterized as
\begin{align}
f_{+}^{D\to \pi}(q^{2})=&\frac{f^{D\to \pi}(0)+c_{+}^{D\to \pi}(z-z_{0})(1+\frac{z+z_{0}}{2})}{1-P_Vq^{2}},\\
f_{0}^{D\to \pi}(q^{2})=&\frac{f^{D\to \pi}(0)+c_{+}^{D\to \pi}(z-z_{0})(1+\frac{z+z_{0}}{2})}{1-P_Sq^{2}},\\
f_{T}^{D\to \pi}(q^{2})=&\frac{f_{T}^{D\to \pi}(0)+c_{T}^{D\to \pi}(z-z_{0})(1+\frac{z+z_{0}}{2})}{1-P_{T}^{D\to \pi}q^{2}}.
\end{align}
The values of all parameters are collected in Tables.~\ref{fitparametersZseries} and \ref{Tab_Fit_Params_z_series_tensor}.

However, for the form factors of $D_s\to K$, $D \to \eta^{(\prime)}$ and $D_s\to \eta^{(\prime)}$, LQCD results are still unavailable till now, and we have to employ the results of other approaches. In this work, we shall adopt the results \cite{Wu:2006rd} of the light-cone sum rules in the framework of the heavy quark effective field theory. In the calculation, in order to describe the behavior of the form factors in the whole kinematically accessible region, we use the double-pole parametrization:
\begin{eqnarray}\label{doublepole}
F^i(q^{2})=\frac{F^i(0)}{1-a\frac{q^{2}}{m_{D}^{2}}+b\left(\frac{q^{2}}{m_{D}^{2}}\right)^2},
\end{eqnarray}
where $F^i(q^{2})$ can be any of the form factors $f_+$ and $f_0$. It should noted that in this work $\eta$ and $\eta^\prime$ are viewed as the mixing states of $\eta_q$ and $\eta_s$ with the mixing angle $\phi=(39.3\pm1.0)^\circ$, and the possible gluonic contribution has not been considered here. The explicit values of each form factor are presented in Table.~\ref{table:formfacor2}.

\begin{table}[!htb]
\renewcommand{\arraystretch}{1.2}
\begin{center}
\caption{Fit parameters for $f_0$, $f_+$ in the $z$-series expansion \cite{Lubicz:2017syv}.}\label{fitparametersZseries}
\begin{tabular}{ c c c c c c }
 \hline
 Decay &$f(0)$ & $c_+$ & $P_V$ (GeV)$^{-2}$ &$c_0$ & $P_S$ (GeV)$^{-2}$\\
 \hline
\hline
 $D\rightarrow \pi$ & 0.6117 (354) & -1.985 (347) & 0.1314 (127) &-1.188 (256) & 0.0342 (122)\\
 \hline
 $D\rightarrow K$ & 0.7647 (308) & -0.066 (333)  &-& -2.084 (283) & -  \\
\hline\hline
\end{tabular}
\end{center}
\end{table}

\begin{table}[!htb]
\renewcommand{\arraystretch}{1.2}
\begin{center}
\caption{Fit parameters for $f_T$ in the $z$-series expansion \cite{Lubicz:2018rfs}.}\label{Tab_Fit_Params_z_series_tensor}
\begin{tabular}{ c c c c }
 \hline
 Decay &$f_T(0)$ & $c_T$ & $P_T$ (GeV)$^{-2}$ \\
 \hline
 \hline
 $D\rightarrow \pi$ & 0.5063 (786) & -1.10 (1.03) & 0.1461 (681)\\
  \hline
 $D\rightarrow K$ & 0.6871 (542) & -2.86 (1.46)  &0.0854 (671) \\
\hline\hline
\end{tabular}
\end{center}
\end{table}

\begin{table}[!htb]
\renewcommand{\arraystretch}{1.2}
\begin{center}
\caption{Form factors of $D_s\to K$, $D\to \eta_q$ and  $D\to \eta_s$  \cite{Wu:2006rd}.}\label{table:formfacor2}
\begin{tabular}{c| c c c c}\hline\hline
Decay & \multicolumn{2}{ c }{$F(0)$} & $a_F$ & $b_F$ \\
\hline
\multirow{2}{*}{$D_s \rightarrow K$}
 & $f_+$ & $0.82^{+0.08}_{-0.07}$ & $1.11^{-0.04}_{+0.07}$ & $0.49^{-0.05}_{+0.06}$ \\
 & $f_0$ & $0.82^{+0.08}_{-0.07}$ & $0.53^{-0.03}_{+0.04}$ & $-0.07^{-0.04}_{+0.04}$ \\
\hline
\multirow{2}{*}{$D \rightarrow \eta_q$}
 & $f_+$ & $0.56^{+0.06}_{-0.05}$ & $1.25^{-0.04}_{+0.05}$ & $0.42^{-0.06}_{+0.05}$ \\
 & $f_0$ & $0.56^{+0.06}_{-0.05}$ & $0.65^{-0.01}_{+0.02}$ & $-0.22^{-0.03}_{+0.02}$ \\
\hline
\multirow{2}{*}{$D_s \rightarrow \eta_s$}
 & $f_+$ & $0.61^{+0.06}_{-0.05}$ & $1.20^{-0.02}_{+0.03}$ & $0.38^{-0.01}_{+0.01}$ \\
 & $f_0$ & $0.61^{+0.06}_{-0.05}$ & $0.64^{-0.01}_{+0.02}$ & $-0.18^{+0.04}_{-0.03}$ \\
\hline\hline
\end{tabular}
\end{center}
\end{table}

The hadronic matrix elements of the vector, scalar and tensor currents between $D$ and $V$ ($V= K^{*}$, $\phi$, $\rho$ and $\omega$) can also be parameterized in terms of eight form factors, respectively,
\begin{align}
\langle V(p_{2},\varepsilon^*)|\bar q\gamma^{\mu}c|D (p_{1})\rangle=&\frac{-2iV(q^2)}{m_{D}+m_{V}}\epsilon_{\mu\nu\alpha\beta}\varepsilon^{*\nu}p_{1}^{\alpha}p_{2}^{\beta},\\
\langle V(p_{2},\varepsilon^*)|\bar q\gamma^{\mu}\gamma_{5} c|D(p_{1})\rangle=&-(m_{D}+m_{V})\varepsilon^{*\mu}A_{1}(q^2)+\frac{\varepsilon^{*}\cdot q}{m_{D }+m_{V}}(p_{1}+p_{2})^{\mu}A_{2}(q^2)\nonumber\\
&\hspace{0.1cm}+2m_{V}\frac{\varepsilon^{*}\cdot q}{q^{2}}q^{\mu}\left(A_{3}(q^2)-A_{0}(q^2)\right),\\
\langle V(p_{2},\varepsilon^*)|\bar q\sigma^{\mu\nu}c|D (p_{1})\rangle=&\epsilon^{\mu\nu\rho\sigma}\Big[\varepsilon_{\rho}^{*}(p_{1}+p_{2})_{\sigma}T_{1}(q^2)+
\varepsilon_{\rho}^{*}q_{\sigma}\frac{m_{D}^{2}-m_{V}^{2}}{q^{2}}(T_{2}(q^2)-T_{1}(q^2))\nonumber\\
&\hspace{0.1cm}+2\frac{\varepsilon^{*}\cdot q}{q^{2}}p_{1\rho}p_{2\sigma}(T_{2}(q^2)-T_{1}(q^2)+\frac{q^{2}}{m_{D}^{2}-m_{V}^{2}}T_{3}(q^2))\Big],
\end{align}
where $A_{0}$ is the abbreviation for
\begin{align}
A_{0}(q^2)=\frac{1}{2m_{V}}\Big[(m_{D}+m_{V})A_{1}(q^2)-(m_{D}-m_{V})A_{2}(q^2)
-\frac{q^2}{m_{D}+m_{V}}A_{3}(q^2)\Big].
\end{align}

In the literatures, there are many studies of these form facators based on different approaches, such as QCD sum rules \cite{Ball:1993tp}, light-cone sum rules (LCSR) \cite{Wu:2006rd,Fu:2018yin, Fu:2020vqd}, quark models \cite{Melikhov:2000yu, Soni:2018adu, Cheng:2017pcq, Faustov:2019mqr, Dai:2018vzz, Chang:2020wvs, Chang:2018zjq}, the covariant light-front quark models \cite{Verma:2011yw,Chang:2019mmh} and LQCD \cite{Bowler:1994zr, Donald:2013pea}. The results of $D \to K^*, \rho$ of LQCD had been released in as early as 1995 \cite{Bowler:1994zr}, however the predicted $D^+ \to  K^{*+}\ell \nu_\ell$ are much larger than the upper limits of experimental results \cite{Fleischer:2019wlx}. The recent undated results are still absent till now. In 2013, the HPQCD collaboration calculated the complete set of axial and vector form factors of $D_s^+ \to \phi$ \cite{Donald:2013pea}, but the ratios at maximum recoil of $A_2(0)/A_1(0)$ and $V(0)/A_1(0)$ are smaller than the experimental data \cite{Soni:2018adu}. In addition, the form factors $D_s^+ \to K^{*}$ and $D \to \omega$ have not been explored till now. Although most results \cite{Verma:2011yw} of the covariant light-front quark model agree well with the experimental data \cite{Zyla:2020zbs} with certain uncertainties, the predicted branching fraction of $D \to K^*\mu^+\nu_\mu$ is also much larger than the experimental data, which lowers its prediction power. As for other results based on quark models, a complete study involving all $D \to V$ processes of all possible currents are in absent, to our best knowledge. For consistency, we shall adopt the results with the LCSR calculation of Ref. \cite{Wu:2006rd}, which is based on the framework of heavy quark effective field theory. This work dates back to 2006, but there exists a more recent determinations \cite{Fu:2018yin, Fu:2020vqd}. In our calculations, the double-pole parametrization as eq.(\ref{doublepole}) is also chosen to interpolate the calculated values of the form factors, and here $F^i$ being any of the form factors $A_1$, $A_2$, $A_3$ and $V$. The values of the parameters are collected in Table.~\ref{tab:form factorv}. In the heavy quark effective theory, the tensor form factors of $D \to V$ are related to the vector and scalar form factors $A_1$, $A_2$, $A_3$ and $V$, and the relations are given as
\begin{align}
T_1(q^2) &= \frac{m_D^2 - m_V^2 + q^2}{2 m_D} \frac{V(q^2)}{m_D + m_V} + \frac{m_D + m_V}{2 m_D} A_1(q^2) ,\\
T_2(q^2) &= \frac{2}{m_D^2 - m_V^2} \Bigg[ \frac{(m_D - y)(m_D + m_V)}{2} A_1(q^2) + \frac{m_D (y^2 - m_V^2)}{m_D + m_V} V(q^2) \Bigg] ,\\
T_3(q^2) &= -\frac{m_D + m_V}{2 m_D} A_1(q^2) + \frac{m_D - m_V}{2 m_D} \big[ A_2(q^2) - A_3(q^2)\big] + \frac{m_D^2 + 3 m_V^2 - q^2}{2 m_D (m_D + m_V)} V(q^2),
\end{align}
where the energy $y$ of the final vector meson is given by
\begin{eqnarray}
y= \frac{m_D^2 + m_V^2 - q^2}{2 m_D}.
\end{eqnarray}

\begin{table}[!htb]
\renewcommand\arraystretch{1.2}
\begin{center}
\caption{\label{tab:form factorv} Form factor of $D(D_{s})\to V$ transitions obtained in the LCSR \cite{Wu:2006rd}.}
\begin{tabular}[h]{c|c|c|c}
\hline
\hline
 &$\hspace{1.15cm}F\hspace{1.15cm}$& $\hspace{1.15cm}a\hspace{1.15cm}$ &$\hspace{1.15cm}b\hspace{1.15cm}$ \\
\hline
\hline
 $A_{1}^{D \to K^{*}}$ & $0.571^{+0.020}_{-0.022}$ & $0.65^{-0.06}_{+0.10}$&$0.66^{-0.18}_{+0.21}$  \\
 $A_{2}^{D \to K^{*}}$ & $0.345^{+0.034}_{-0.037}$ & $1.86^{+0.05}_{-0.22}$&$-0.91^{+0.48}_{-0.97}$  \\
 $A_{3}^{D \to K^{*}}$ & $-0.723^{+0.065}_{-0.077}$ & $1.32^{+0.14}_{-0.09}$&$1.28^{+0.22}_{-0.21}$ \\
 $V^{D \to K^{*}}$ & $0.791^{+0.024}_{-0.026}$ & $1.04^{-0.17}_{+0.25}$&$2.21^{-0.12}_{+0.37}$ \\
 \hline
 $A_{1}^{D_{s} \to K^{*}}$ & $0.589^{+0.040}_{-0.042}$ & $0.56^{-0.02}_{+0.02}$&$-0.12^{+0.03}_{-0.02}$  \\
 $A_{2}^{D_{s} \to K^{*}}$ & $0.315^{+0.024}_{-0.018}$ & $0.15^{+0.22}_{-0.14}$&$0.24^{-0.94}_{+0.83}$  \\
 $A_{3}^{D_{s} \to K^{*}}$ & $-0.675^{+0.027}_{-0.037}$ & $0.48^{-0.11}_{+0.13}$&$-0.14^{+0.18}_{-0.17}$ \\
 $V^{D_{s} \to K^{*}}$ & $0.771^{+0.049}_{-0.049}$ & $1.08^{-0.02}_{+0.02}$&$0.13^{-0.03}_{-0.02}$ \\
 \hline
 $A_{1}^{D_{s} \to \phi}$ & $0.569^{+0.046}_{-0.049}$ & $0.84^{-0.05}_{+0.06}$&$0.16^{-0.01}_{+0.01}$  \\
 $A_{2}^{D_{s} \to \phi}$ & $0.304^{+0.021}_{-0.017}$ & $0.24^{+0.18}_{-0.05}$&$1.25^{-1.08}_{+1.02}$ \\
 $A_{3}^{D_{s} \to \phi}$ & $-0.757^{+0.029}_{-0.039}$ & $0.60^{-0.02}_{+0.07}$&$0.60^{+0.31}_{-0.33}$ \\
 $V^{D_{s} \to \phi}$ & $0.778^{+0.057}_{-0.062}$ & $1.37^{-0.05}_{+0.04}$&$0.52^{+0.04}_{-0.06}$ \\
  \hline
 $A_{1}^{D\to \rho}$ & $0.599^{+0.035}_{-0.030}$ & $0.44^{-0.06}_{+0.10}$&$0.58^{-0.04}_{+0.23}$  \\
 $A_{2}^{D\to \rho}$ & $0.372^{+0.026}_{-0.031}$ & $1.64^{-0.16}_{+0.10}$&$0.56^{-0.28}_{+0.04}$ \\
 $A_{3}^{D\to \rho}$ & $-0.719^{+0.055}_{-0.066}$ & $1.05^{+0.15}_{-0.15}$&$1.77^{-0.11}_{+0.20}$ \\
 $V^{ D\to \rho}$ & $0.801^{+0.044}_{-0.036}$ & $0.78^{-0.20}_{+0.24}$&$2.61^{+0.29}_{-0.04}$ \\
 \hline
 $A_{1}^{D\to \omega }$ & $0.556^{+0.033}_{-0.028}$ & $0.45^{-0.05}_{+0.09}$&$0.54^{-0.10}_{+0.17}$  \\
 $A_{2}^{D\to \omega }$ & $0.333^{+0.026}_{-0.030}$ & $1.67^{-0.15}_{+0.09}$&$0.44^{-0.29}_{+0.05}$ \\
 $A_{3}^{D\to \omega }$ & $-0.657^{+0.053}_{-0.063}$ & $1.07^{+0.17}_{-0.14}$&$1.77^{+0.14}_{+0.07}$ \\
 $V^{ D\to \omega }$ & $0.742^{+0.041}_{-0.034}$ & $0.79^{-0.20}_{+0.22}$&$2.52^{+0.28}_{-0.13}$ \\
\hline
\hline
\end{tabular}
\end{center}
\end{table}

\subsection{The Helicity Amplitudes}
In SM, the transition $c\to s\ell^+ \nu_\ell$ can be viewed as $c\to sW^{*+}$, and the off-shell $W^{*+}$ decays to $\ell^+ \nu_\ell$, subsequently. It is known to us that the off-shell $W^{*+}$ has four helicities, namely $\lambda_{W}=\pm1,0$ ($J=1$) and $\lambda_{W}=0$ ($J=0$), and only the $W^{*+}$ boson has a timelike polarization, with $J=1,0$ denoting the two angular momenta in the rest frame of the $W^{*}$ boson. In order to distinguish the two $\lambda_{W}=0$ states we set $\lambda_{W}=0$ for $J=1$ and $\lambda_{W}=t$ for $J=0$. In the $D$ meson rest frame, we set the $z$-axis to be along the moving direction of $W^{*+}$, and write the polarization vectors of the $W^{*+}$ as
\begin{align}
\epsilon^{\mu}(\pm)=\frac{1}{2}(0,1,\mp i,0),\,\,\,\epsilon^{\mu}(0)=-\frac{1}{\sqrt{q^{2}}}(q_{3},0,0,q_{0}),\,\,\,\epsilon^{\mu}(t)=-\frac{q^{\mu}}{\sqrt{q^{2}}},
\end{align}
where $q^{\mu}$ is the four-momentum of the $W^{*+}$. The polarization vectors of the virtual $W^{*+}$ satisfy the orthogonality and completeness relations
\begin{align}
\epsilon^{*\mu}(m)\epsilon_{\mu}(n)=g_{mn},\,\,\,\sum_{m,n}\epsilon^{*\mu}(m)\epsilon^{\nu}(n)g_{mn}=g^{\mu\nu},
\end{align}
where $g_{mn}$ is diag($+,-,-,-$) for $m,n=t,\pm,0$.

In the calculations, the total matrix can be factorized into lepton part and hadron part, both of which are not the Lorentz invariant. When inserting the completeness relations of $W^{*+}$, both the hadron side and the lepton side become Lorentz invariant, which make us choose the coordinate system arbitrarily. Thereby, the hadron side can be analyzed in the initial state $D$ meson rest frame, and the lepton side is analyzed in the virtual $W^{*+}$ rest frame. We then calculate the helicity amplitudes of $D\to PW^{*+}$ transition as
\begin{eqnarray}\label{eq:K Helicity}
&&H^{PV}_{\lambda_{W}}(q^{2})=\epsilon^{*}_{\mu}(\lambda_{W})\langle P(p_{2})|\bar q\gamma^{\mu}c|D(p_{1})\rangle\\
&&H^{PS}(q^{2})=\langle P(p_{2})|\bar qc|D(p_{1})\rangle, \\
&&H^{PT}_{\lambda_{W},\lambda_{W}^{\prime}}(q^{2})=\epsilon^{*}_{\mu}(\lambda_{W})\epsilon^{*}_{\nu}(\lambda_{W}^{\prime})
\langle P(p_{2})|\bar q\sigma^{\mu\nu}c|D(p_{1})\rangle.
\end{eqnarray}
Similarly, the helicity amplitudes of the $D\to VW^{*+}$transition are given as
\begin{eqnarray}\label{eq:M Helicity}
&&H^{VAL(VAR)}_{\lambda_{W},\varepsilon_{V}}(q^{2})=\epsilon^{*}_{\mu}(\lambda_{W})
\langle V(p_{2},\varepsilon^*)|\bar q\gamma^{\mu}(1\pm\gamma^{5})c|D_{(s)}(p_{1})\rangle\nonumber\\
&&H^{SPL(SPR)}_{\varepsilon_{V}}(q^{2})=\langle V(p_{2},\varepsilon^*)|\bar q\gamma^{\mu}(1\pm\gamma^{5})c|D_{(s)}(p_{1})\rangle,\nonumber\\
&&H^{T}_{\lambda_{W},\lambda_{W}^{'},\varepsilon_{V}}(q^{2})=\epsilon^{*}_{\mu}(\lambda_{W})\epsilon^{*}_{\nu}(\lambda_{W}^{'})\langle V(p_{2},\varepsilon^*)|\bar q\sigma^{\mu\nu}(1-\gamma^{5})c|D_{(s)}(p_{1})\rangle.
\end{eqnarray}

For an arbitrary decay $D\to F \ell^+\nu_\ell$ ($F=P,V$), due to the conservation of helicity, $-\varepsilon_{F} + \lambda_{W^*}=0$ is satisfied, $\varepsilon_{F}$ being the polarization the final state $F$. So, only five helicity amplitudes contribute to $D\to P\ell^+\nu_\ell$ decays, which are given as
\begin{align}\label{eq:SM hadronic amplitudes}
H^{PV}_{0}=\frac{f_{+}\sqrt{Q_{+}Q_{-}}}{\sqrt{q^2}}\,,
H^{PV}_{t}=&\frac{f_{0}M_{+}M_{-}}{\sqrt{q^2}}\,,
H^{PS}=\frac{f_{0}M_{+}M_{-}}{m_{c}-m_{q}}\,,\nonumber\\
H^{PT}_{0,t}=-\frac{f_{T}\sqrt{Q_{+}Q_{-}}}{M_{+}}&\,,
H^{PT}_{1,-1}=-\frac{f_{T}\sqrt{Q_{+}Q_{-}}}{M_{+}}\,.
\end{align}
Because the vector meson are polarized, the helicity amplitudes that contribute to the $D \to V\ell^+\nu_\ell$ decay are presented as:
\begin{eqnarray}
&&H^{V}_{1}\equiv H^{VAL}_{1,1}=-H^{VAR}_{-1,-1}=A_{1}M_{+}-\frac{\sqrt{Q_{+}Q_{-}}V}{M_{+}}\,,\nonumber\\
&&H^{V}_{-1}\equiv H^{VAL}_{-1,-1}=-H^{VAR}_{1,1}=A_{1}M_{+}+\frac{\sqrt{Q_{+}Q_{-}}V}{M_{+}}\,,\nonumber\\
&&H^{V}_{0}\equiv H^{VAL}_{0,0}=-H^{VAR}_{0,0}=-\frac{A_{1}M_{+}^{2}(M_{+}M_{-}-q^{2})-A_{2}Q_{+}Q_{-}}{2m_{F}M_{+}\sqrt{q^2}}\,,\nonumber\\
&&H^{V}_{t}\equiv H^{VAL}_{t,0}=-H^{VAR}_{t,0}=-A_{0}\sqrt{\frac{Q_{+}Q_{-}}{q^{2}}}\,,\nonumber\\
&&H^{S}\equiv H^{SPL}_{0}=-H^{SPR}_{0}=\frac{A_{0}\sqrt{Q_{+}Q_{-}}}{m_{c}+m_{q}}\,,\nonumber\\
&&H^{T}_{+}\equiv H^{T}_{1,t,1}=H^{T}_{1,0,1}=\frac{\sqrt{Q_{+}Q_{-}}T_{1}+M_{+}M_{-}T_{2}}{\sqrt{q^2}}\,,\nonumber\\
&&H^{T}_{-}\equiv -H^{T}_{-1,t,-1}=H^{T}_{-1,0,-1}=\frac{\sqrt{Q_{+}Q_{-}}T_{1}-M_{+}M_{-}T_{2}}{\sqrt{q^2}}\,,\nonumber\\
&&H^{T}_{0}\equiv H^{T}_{1,-1,0}=H^{T}_{0,t,0}=\frac{-M_{+}M_{-}(m_{D}^{2}+3m_{F}^{2}-q^{2})T_{2}+Q_{+}Q_{-}T_{3}}{2M_{+}M_{-}m_{F}}\,,
\end{eqnarray}
where $M_{\pm}=m_{D}\pm m_{F}$ and $Q_{\pm}=M_{\pm}^{2}-q^{2}$, respectively. In the above equations, $m_{F}$ is the mass of the final state meson. Due to the absence of complicated QCD, the leptonic amplitudes can be calculated directly, and the explicit results can be found in \cite{Ivanov:2016qtw}.

\subsection{Observables}
With the hadronic helicity amplitudes and the leptonic amplitudes, we then write down the two-fold differential angular decay distribution of $D\to P\ell^+\nu_\ell$ decay as
\begin{equation}\label{eq:P-differential angular}
\frac{{d}^2\Gamma(D\to P\ell^+\nu_\ell)}{{ d}q^2\,{d}\cos\theta_\ell}
=\frac{G_F^2|V_{cq}|^2\sqrt{Q_{+}Q_{-}}}{256\pi^3m_{D}^3}\left(1-\frac{m_\ell^2}{q^2}\right)^2\left[q^2A_1^{P}+\sqrt{q^2}m_{l}A_{2}^{P}
+m_{l}^{2}A_3^{P}\right]\,,
\end{equation}
with
\begin{multline}
A_1^{P}=|C_{SL}+C_{SR}|^2|H^{PS}|^{2}+{\rm Re}\Big[(C_{SL}+C_{SR})\,C_{T}^{*}\Big]H^{PS}(H^{PT}_{0,t}+H^{PT}_{0,-1})\cos\theta_\ell \\
 +4|C_{T}|^2|H^{PT}_{0,t}+H^{PT}_{1,-1}|^{2}\cos^{2}\theta_\ell+|1+C_{VL}+C_{VR}|^2|H^{PV}_{0}|^{2}\sin^{2}\theta_\ell\,,
\end{multline}
\begin{multline}
A_{2}^{P}=2\Big\{{\rm Re}\Big[(C_{SL}+C_{SR})\,(1+C_{VL}+C_{VR})^{*}\Big]H^{PS}H^{PV}_{t} \\
 -2{\rm Re}\Big[C_{T}\,(1+C_{VL}+C_{VR})^{*}\Big]H^{PV}_{0}(H^{PT}_{0,t}+H^{PT}_{1,-1})\Big\} \\
 -2\Big\{{\rm Re}\Big[(C_{SL}+C_{SR})\,(1+C_{VL}+C_{VR})^{*}\Big]H^{PS}H^{PV}_{0}\\
 -2{\rm Re}\Big[C_{T}\,(1+C_{VL}+C_{VR})^{*}\Big]H^{PV}_{t}(H^{PT}_{0,t}+H^{PT}_{1,-1})\Big\}\cos\theta_\ell,
\end{multline}
\begin{multline}
A_3^{P}=4|C_{T}|^{2}|H^{PT}_{0,t}+H^{PT}_{1,-1}|^{2}\sin^{2}\theta_\ell \\
 +|1+C_{VL}+C_{VR}|^{2}(|H^{PV}_{0}|^{2}\cos^{2}\theta_\ell-2H^{PV}_{0}H^{PV}_{T}\cos\theta_\ell+|H^{PV}_{t}|^{2})\,,
\end{multline}
with $\theta_\ell$ being the angle between the charged lepton and the opposite direction of the motion of the final meson in the virtual $W^{*+}$ rest frame.

For $D \to V\ell^+\nu_\ell$ decay, we also have the similar result
\begin{equation}\label{eq:V-differential angular}
\frac{{ d}^2\Gamma(D\to V\ell^+\nu_\ell)}{{ d}q^2\,{d}\cos\theta_\ell}
=\frac{G_F^2|V_{cq}|^2\sqrt{Q_{+}Q_{-}}}{512\pi^3m_{D}^3}\left(1-\frac{m_\ell^2}{q^2}\right)^2
\left[q^2A_1^{V}+4\sqrt{q^2}m_{l}A_2^{V}+m_{l}^{2}A_3^{V}\right]\,,
\end{equation}
with
\begin{multline}\label{eq:P-coefficients}
A_1^{V}=|1+C_{VL}|^2\Big[(1+\cos\theta_\ell)^{2}|H^{V}_{-1}|^{2}+2\sin^{2}\theta_\ell |H^{V}_{0}|^{2}+(1-\cos\theta_\ell)^{2}|H^{V}_{1}|^{2}\Big] \\
+|C_{VR}|^2\Big[(1-\cos\theta_\ell)^{2}|H^{V}_{-1}|^{2}+2\sin^{2}\theta_\ell |H^{V}_{0}|^{2}+(1+\cos\theta_\ell)^{2}|H^{V}_{1}|^{2}\Big] \\
+16|C_{T}|^2\Big[\sin^{2}\theta_\ell |H^{T}_{-1}|^{2}+|H^{T}_{1}|^{2}+\cos^{2}\theta_\ell(2|H^{T}_{0}|^{2}-|H^{T}_{1}|^{2})\Big] \\
+2|C_{SL}-C_{SR}|^2|H^{S}|^{2}+16{\rm Re}\Big[(C_{SL}-C_{SR})C_{T}^{*}\Big]\cos\theta_\ell H^{S}H^{T}_{0} \\
-4{\rm Re}\Big[C_{VL}C_{VR}^{*}\Big]\Big[\sin^{2}\theta_\ell |H^{V}_{0}|^{2}+(1+\cos^{2}\theta_\ell)H^{V}_{-1}H^{V}_{1}\Big]\,,
\end{multline}
\begin{multline}
A_{2}^{V}={\rm Re}\Big[(1+C_{VL}-C_{VR})(C_{SL}-C_{SR})^{*}\Big](H^{V}_{t}H^{S}-H^{V}_{0}H^{S}\cos\theta_\ell) \\
-4{\rm Re}\Big[C_{T}C_{VR}^{*}\Big]\Big[(\cos\theta_\ell-1)H^{T}_{1}H^{V}_{-1}-H^{T}_{0}H^{V}_{0}
+(\cos\theta_\ell+1)H^{T}_{-1}H^{V}_{1}+\cos\theta_\ell H^{T}_{0}H^{V}_{t}\Big] \\
+4{\rm Re}\Big[C_{T}(1+C_{VL})^{*}\Big]\Big[(\cos\theta_\ell+1)H^{T}_{-1}H^{V}_{-1}-H^{T}_{0}H^{V}_{0}
+(\cos\theta_\ell-1)H^{T}_{1}H^{V}_{1}+\cos\theta_\ell H^{T}_{0}H^{V}_{t}\Big],
\end{multline}
\begin{multline}
A_3^{V}=16|C_{T}|^{2}\Big[(1+\cos\theta_\ell)^{2}|H^{T}_{-1}|^{2}+2\sin^{2}\theta_\ell|H^{T}_{0}|^{2}
+(1-\cos\theta_\ell)^{2}|H^{T}_{1}|^{2}\Big] \\
+(|1+C_{VL}|^{2}+|C_{VR}|^{2})\Big[(1-\cos^{2}\theta_\ell)(|H^{V}_{-1}|^{2}+|H^{V}_{1}|^{2})\Big] \\
+|1+C_{VL}-C_{VR}|^{2}\Big[\cos^{2}\theta_\ell|H^{V}_{0}|^{2}+|H^{V}_{t}|^{2}-2\cos\theta_\ell H^{V}_{0}H^{V}_{t}\Big] \\
-4{\rm Re}\Big[(1+C_{VL})C_{VR}^{*}\Big]\sin^{2}\theta_\ell H^{V}_{1}H^{V}_{-1}\,.
\end{multline}

After integrating out the $\cos\theta_\ell$ in eqs.(\ref{eq:P-differential angular}) and (\ref{eq:V-differential angular}), we obtain the differential decay rate $ d\Gamma(D\to F\ell^+\nu_\ell)/dq^{2}$. Then, the total branching fraction can be given as
\begin{align}
\mathcal{B}(D\to F\ell^+\nu_\ell)=\tau_{D}\int^{M^{2}_{-}}_{m^{2}_{l}}dq^{2}\frac{d\Gamma(D\to F\ell^+\nu_\ell)}{dq^{2}},
\end{align}
with $\tau_{D}$ being the lifetime of the $D$ meson.

Besides the branching fractions, we also focus on other observables. We define the forward-backward asymmetry in the lepton-side as
\begin{eqnarray}
A_{FB}(q^{2})=\frac{\int^{1}_{0} d \cos\theta_\ell \frac{d^2\Gamma}{dq^2d\cos\theta_\ell}-\int^{0}_{-1}d\cos\theta_\ell \frac{d^2\Gamma}{dq^2d\cos\theta_\ell}}
{\int^{1}_{0} d \cos\theta_\ell \frac{d^2\Gamma}{dq^2d\cos\theta_\ell}+\int^{0}_{-1}d\cos\theta_\ell \frac{d^2\Gamma}{dq^2d\cos\theta_\ell}}.
\end{eqnarray}
Similar to $B \to D^{*}\ell \bar \nu_{\ell}$ \cite{Sakaki:2014sea}, the $q^{2}$-dependent longitudinal polarization of the vector meson can be defined as
\begin{eqnarray}
P_L^{V}(q^2)=\frac{d\Gamma (\epsilon_V=0)/dq^2}{d\Gamma/dq^2},
\end{eqnarray}
and the lepton polarization parameter can also be defined as
\begin{eqnarray}
P_F^{\ell}(q^2)=\frac{\frac{d\Gamma(\lambda_{\ell}=1/2)}{dq^2}- \frac{d\Gamma(\lambda_{\ell}=-1/2)}{dq^2}}{\frac{d\Gamma(\lambda_{\ell}=1/2)}{dq^2}+ \frac{d\Gamma(\lambda_{\ell}=-1/2)}{{dq^2}}}.
\end{eqnarray}

\section{Parameters} \label{sec:paramerter}
When searching for the effects of NP in $D$ decays, we should be careful on the values of CKM matrix elements $|V_{cd}|$ and $|V_{cs}|$. Generally, these two CKM matrix elements could be extracted from leptonic or semileptonic $D$ decays, assuming that SM is correct, such as discussions in \cite{Riggio:2017zwh}. In the current work, it is paradoxical to use CKM matrix elements extracted from the above experiments when investigating these decays in the presence of NP contributions. In order to avoid such contradiction, we therefore adopt the values as \cite{Fleischer:2019wlx}
\begin{eqnarray}
|V_{cd}|=0.2242\pm0.0005, \hspace{1cm}|V_{cs}|=0.9736\pm0.0001,
\end{eqnarray}
which are obtained by using the unitary property of CKM matrix elements and combining the measurements of $B$ decays and mixings. The strategy was presented in detail in Ref.~\cite{Fleischer:2019wlx}. Similarly, the decay constants $f_D$ are usually determined by the pure leptonic decays of $D$ mesons. In order to avoid the similar paradox, we use the results from the LQCD \cite{Aoki:2019cca} directly, and they are given as
\begin{eqnarray}
f_{D^+} = (209.0 \pm 2.4) \text{ MeV}, \hspace{1cm} f_{D^+_s} = (248.0 \pm 1.6)  \text{ MeV}.
\end{eqnarray}
The other parameters, such as the Fermi constant, masses of mesons and lifetimes of $D$ mesons, are taken from PDG \cite{Zyla:2020zbs}

\section{Leptonic Decays} \label{sec:leptonicdecays}
Besides the semi-leptonic decays, the effective Lagrangian also controls the pure leptonic decay $D \to \ell^+\nu_\ell$ , the formula for the branching fraction of which is given by
\begin{eqnarray}\label{pld}
\mathcal B(D^+ \to \ell^+\nu_\ell) &=&\tau_{D} \frac{G_F^2}{  8\pi}  |V_{cq}|^2f_{D}^2m_{D} m_\ell^2 \left( 1 -\frac{m_\ell^2}{ m_{D}^2} \right)^2
\nonumber
\\&&\times \left| 1+C^\ell_{VL}-C^\ell_{VR} + \frac{m_{D}^2}{m_\ell(m_c+m_q)} (C^\ell_{SR}-C^\ell_{SL}) \right|^2 (1+\delta^{\ell}_{em}) \,.
\end{eqnarray}
Because the tensor operator $O_T$ is antisymmetric in indexes $\mu$ and $\nu$,  it cannot contribute to the pure leptonic decays. $\delta^{\ell}_{em}\sim (0-3)\%$ is from the electromagnetic corrections \cite{Becirevic:2009aq}, the study of which is out of the scope of this work, we will not discuss it here. It should be emphasized that the pure leptonic decays of $D$ mesons should be helicity suppressed, which is reflected by the proportionality of the branching fractions to $m_\ell^2$. As a results, the branching fractions of decay modes involving the electron are too small to have been measured till now.

\begin{table}[!htb]
\begin{center}
\caption{Branching ratios of leptonic $D_{s}^+$ decays calculated in the SM and comparison with the currently available
experimental values.}\label{leptonicresult}
\renewcommand{\arraystretch}{1.2}
\begin{tabular}{ cccc }
 \hline
 Decay & SM & Experiment  \\
 \hline
 \hline
 $\mathcal{B}(D^+ \rightarrow e^+ \nu_{e}) $ 		& $(9.17 \pm 0.22) \times 10^{-9}  $ & $<  8.8 \times 10^{-6}$ & \cite{Zyla:2020zbs} \\
 $\mathcal{B}(D^+ \rightarrow \mu^+ \nu_{\mu})$ 	& $(3.89 \pm 0.09) \times 10^{-4}$ & $(3.74 \pm 0.17) \times 10^{-4} $& \cite{Zyla:2020zbs} \\
 $\mathcal{B}(D^+ \rightarrow \tau^+ \nu_{\tau})$	& $(1.04 \pm 0.03) \times 10^{-3}$ & $(1.20 \pm 0.27)\times 10^{-3}$ & \cite{Ablikim:2019rpl}  \\
 \hline
 $\mathcal{B}(D_s^+ \rightarrow e^+ \nu_{e}) $		& $(1.24 \pm 0.02) \times 10^{-7}$ &$<  8.3 \times 10^{-5}  $& \cite{Zyla:2020zbs}\\
 $\mathcal{B}(D_s^+ \rightarrow \mu^+ \nu_{\mu}) $	& $(5.28 \pm 0.08) \times 10^{-3}$& $(5.50 \pm 0.23) \times 10^{-3} $&  \cite{Zyla:2020zbs} \\
 $\mathcal{B}(D_s^+ \rightarrow \tau^+ \nu_{\tau}) $	& $(5.14 \pm 0.08) \times 10^{-2}$& $(5.48 \pm 0.23)\times 10^{-2} $&  \cite{Zyla:2020zbs}\\
 \hline
\end{tabular}
\end{center}

\end{table}

\begin{figure*}[!htb]
\begin{center}
\includegraphics[width=7.0cm]{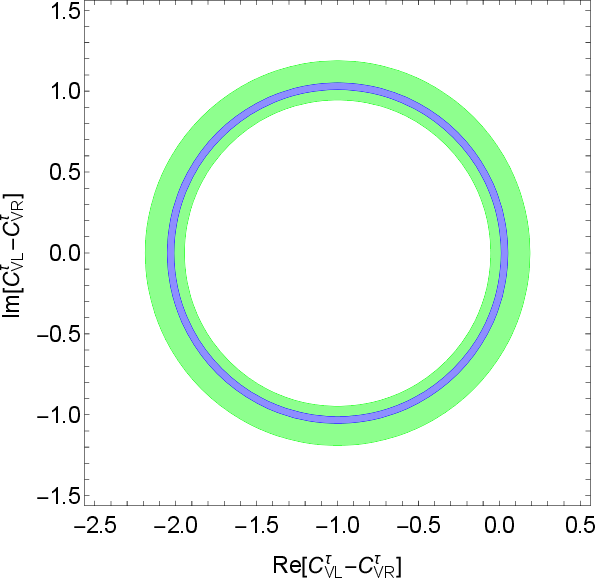}\,\,\,\,
\includegraphics[width=7.0cm]{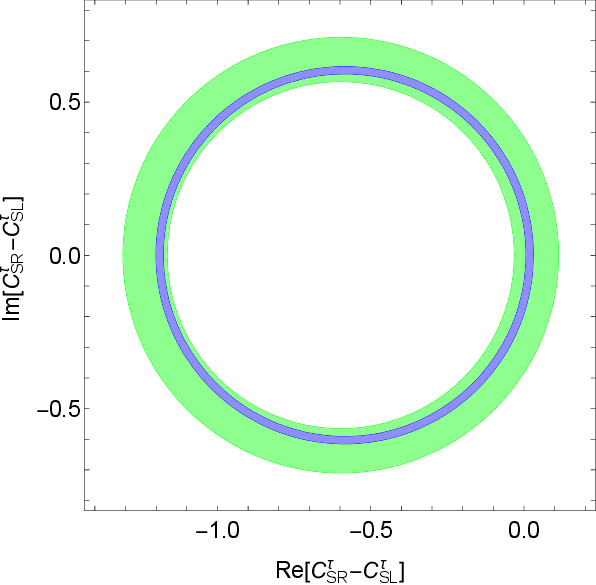}
\caption{Allowed regions of the effective couplings $C_{VL}^\tau-C_{VR}^\tau$ (left panel) and $C_{SR}^\tau-C_{SL}^\tau$ (right panel), extracted from the branching fractions of the decay modes $\mathcal{B}(D^+ \to \tau^+ \nu_{\tau})$ (green) and $\mathcal{B}(D_s^+ \to \tau^+ \nu_{\tau}) $ (blue), respectively.}
\label{fig-1}
\end{center}
\end{figure*}

\begin{figure*}[!htb]
\begin{center}
\includegraphics[width=7.0cm]{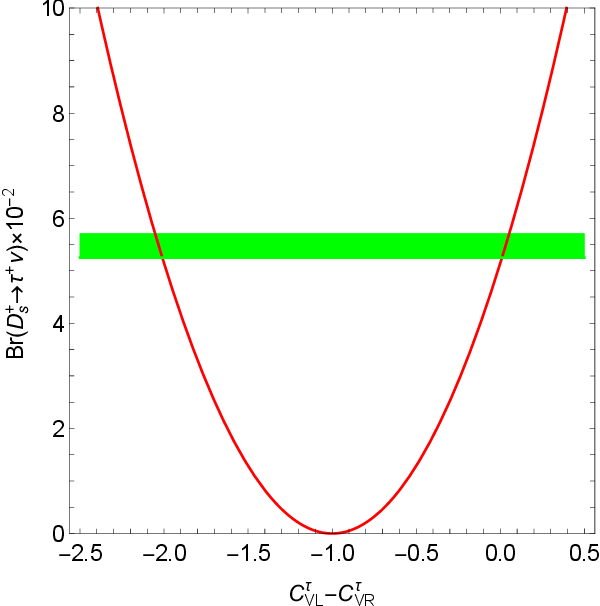}\,\,\,\,\,\,\,\,\,\,\,\,
\includegraphics[width=7.0cm]{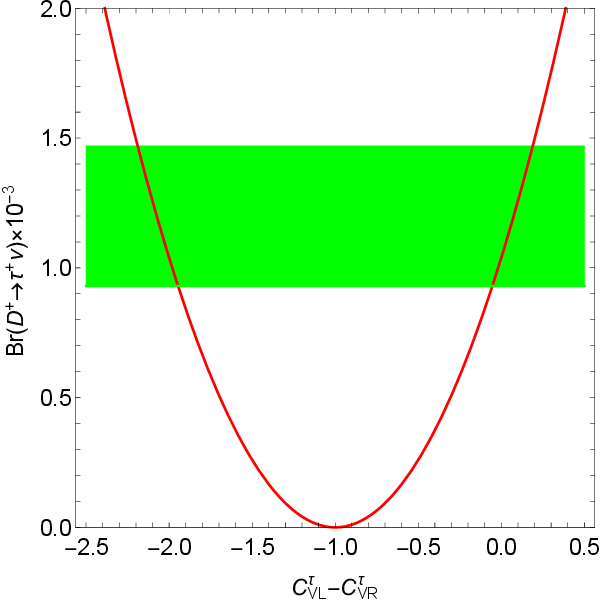}\\
\vspace{1cm}
\includegraphics[width=7.0cm]{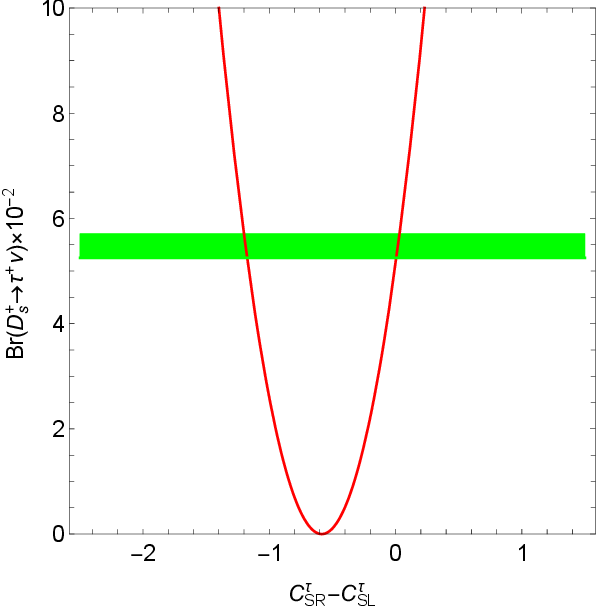}\,\,\,\,\,\,\,\,\,\,\,\,
\includegraphics[width=7.0cm]{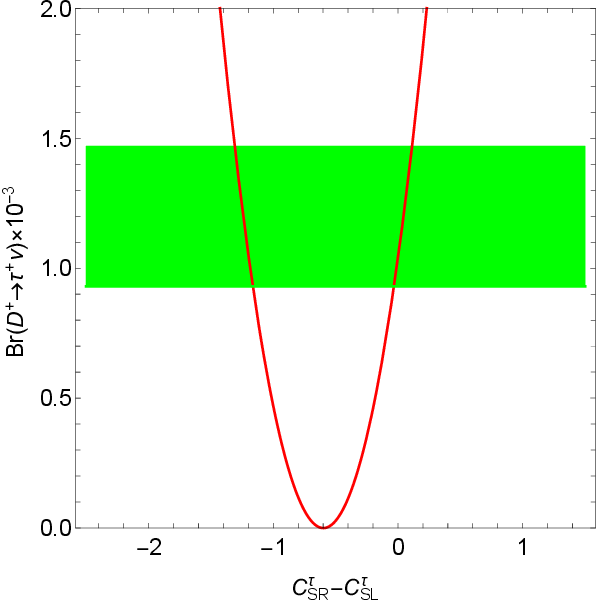}
\caption{The dependence of $\mathcal{B}(D^+ \to \tau^+ \nu_{\tau})$ and $\mathcal{B}(D_s^+ \to \tau^+ \nu_{\tau}) $ on $C_{VL}^\tau-C_{VR}^\tau$ and $C_{SR}^\tau-C_{SL}^\tau$, respectively. The green bands are the experimental data.}
\label{fig-2}
\end{center}
\end{figure*}

\begin{figure*}[!htb]
\begin{center}
\includegraphics[width=7.0cm]{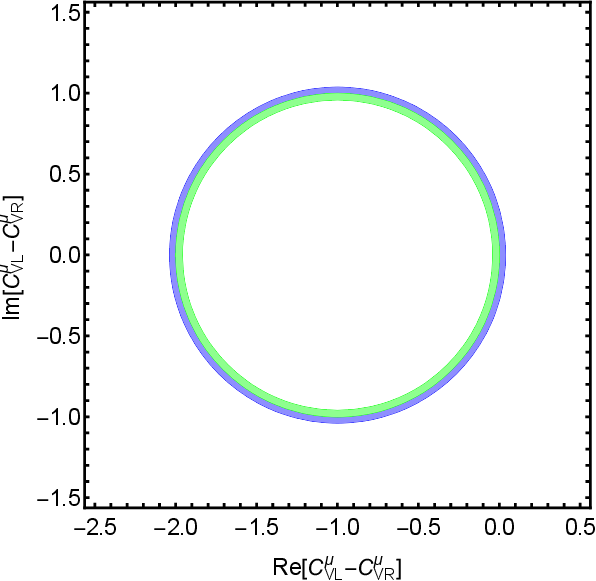}\,\,\,\,\,\,\,\,\,\,\,\,
\includegraphics[width=7.0cm]{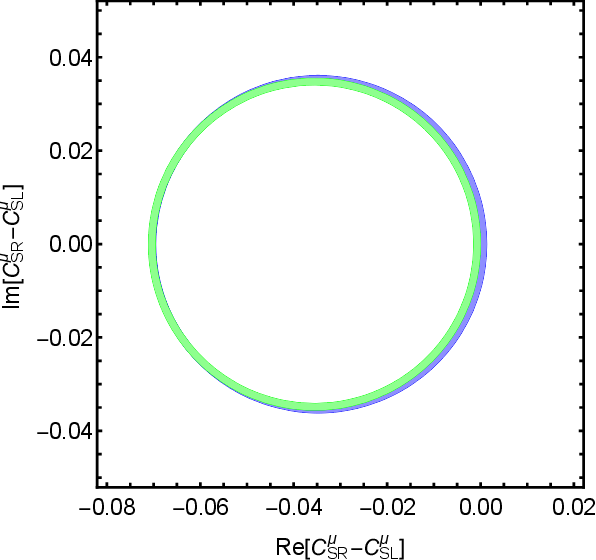}
\caption{Allowed regions of the effective couplings $C_{VL}^\mu-C_{VR}^\mu$ (left panel) and $C_{SR}^\mu-C_{SL}^\mu$ (right panel), extracted from the branching fractions of the decay modes $\mathcal{B}(D^+ \to \mu^+ \nu_{\mu})$ (green) and $\mathcal{B}(D_s^+ \to \mu^+ \nu_{\mu}) $ (blue), respectively.}
\label{fig-3}
\end{center}
\end{figure*}

In Table.~\ref{leptonicresult}, we present the numerical results in SM and the corresponding experimental results, which are same as the results in \cite{Fleischer:2019wlx}. In our calculations, the major uncertainties are from the decay constants of $D$ mesons, and the electromagnetic corrections have not been included here. From the table, it seems that the center theoretical values of $\mathcal{B}(D^+ \to \tau^+ \nu_{\tau})$ and $\mathcal{B}(D_s^+ \to \tau^+ \nu_{\tau}) $ are a bit smaller than the experimental data, which implies that there is room for the NP to survive. If NP affects the transition $c\to q \tau^+ \nu_\tau$, we can constrain the Wilson coefficients with the experimental data. Supposing the Wilson coefficients are complex temporarily, we present the allowed parameter spaces of the $|C_{VL}^\tau-C_{VR}^\tau|$ and $|C_{SR}^\tau-C_{SL}^\tau|$ in Fig.~\ref{fig-1}, where one can find that the major constraint is from the decay $D_s^+ \to \tau^+ \nu_{\tau}$. In Fig.~\ref{fig-2}, we suppose the Wilson coefficients are real and show the dependencies of $\mathcal{B}(D^+ \to \tau^+ \nu_{\tau})$ and $\mathcal{B}(D_s^+ \to \tau^+ \nu_{\tau}) $ on $|C_{VL}^\tau-C_{VR}^\tau|$ and $|C_{SR}^\tau-C_{SL}^\tau|$, respectively. Moreover, the current experimental results are also shown in these two plots. From the $\mathcal{B}(D_s^+ \to \tau^+ \nu_{\tau}) =(5.48 \pm 0.23)\times 10^{-2} $, we have the $|1+C_{VL}^\tau-C_{VR}^\tau|=1.03\pm0.02$, while $|1+C_{VL}^\tau-C_{VR}^\tau|=1.07\pm0.12$ is obtained from $\mathcal{B}(D^+ \to \tau^+ \nu_{\tau}) =(1.20 \pm 0.27)\times 10^{-3}$. These possible allowed ranges can be used to constrain the model in which the new particles can contribute the operators $O_{VL}$ or $O_{VR}$, such as leptoquark models, R-parity violation supersymmetry models, and models with $W^\prime$ or $W_R$.  Also, we obtain the $C_{SR}^\tau-C_{SL}^\tau=-1.19\pm0.02 $ or $ 0.02\pm0.01 $, which are helpful for constraining the models with charged scalars, such as Two-Higgs doublet models, the left-right models and some leptoquark models. For decay modes with final states $\mu^+\nu_\mu$, the theoretical prediction of $\mathcal{B}(D_s^+ \to \mu^+\nu_\mu) $ is a bit smaller than the center value of experimental data, while the situation is reverse for decay $D^+ \to \mu^+\nu_\mu$. Therefore, as shown in Fig.~\ref{fig-3}, there are still allowed parameter regions left for both $|C_{VL}^\mu-C_{VR}^\mu|$ and $|C_{SR}^\mu-C_{SL}^\mu|$ at $1 \sigma$ confidence level.

\begin{figure*}[!htb]
\begin{center}
\includegraphics[width=7.0cm]{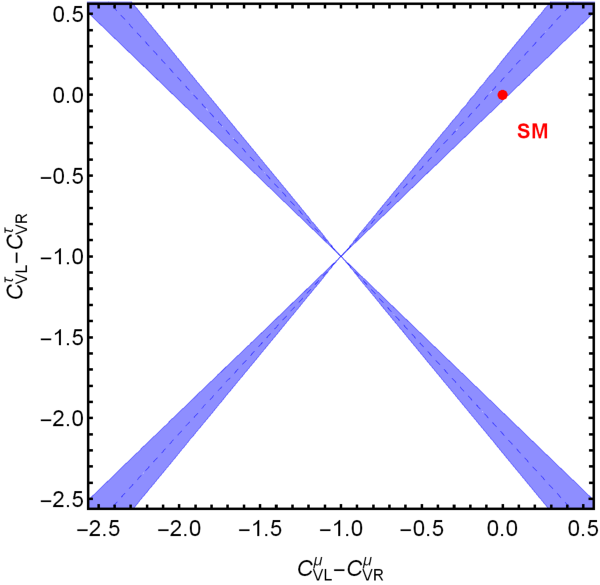}\,\,\,\,\,\,\,\,\,\,\,\,
\includegraphics[width=7.0cm]{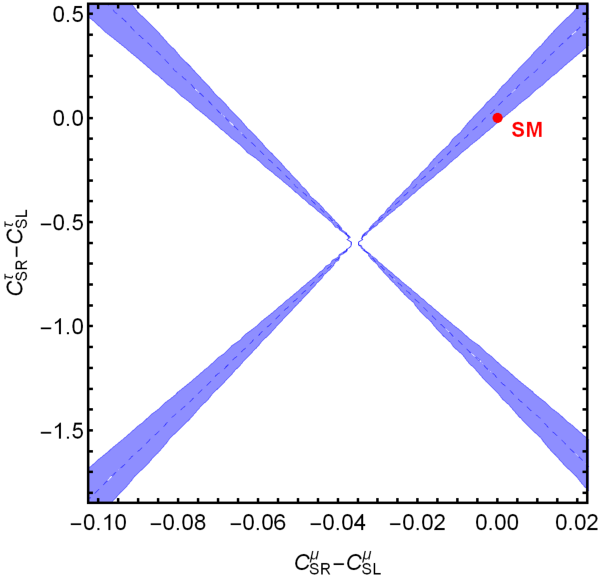}
\caption{The allowed regions in the ($C_{VL}^\mu-C_{VR}^\mu$)-($C_{VL}^\tau-C_{VR}^\tau$)(left) and  ($C_{SL}^\mu-C_{SR}^\mu$)-($C_{SL}^\tau-C_{SR}^\tau$) (right) planes following from the ratio $({\cal R}^{\tau}_{\mu})^D$, respectively. The red dots indicate the SM predictions}
\label{fig-4}
\end{center}
\end{figure*}

In order to test the lepton flavour universality, one can define the ratio of two leptonic decays as \cite{Fleischer:2019wlx}
\begin{eqnarray} \label{ratio}
{\cal R}_{\ell_2}^{\ell_1}=\frac{\mathcal{B}(D_{(s)}^+ \rightarrow \ell_1^+ \nu_{\ell_1})}{\mathcal{B}(D_{(s)}^+ \rightarrow \ell_2^+ \nu_{\ell_2})}.
\end{eqnarray}
This observable is theoretically clean, because both uncertainties from the decay constants and the CKM matrix elements have been cancelled out by each other. Within the available experimental data, the constraints on each coefficient have been discussed in great detail in Ref.\cite{Fleischer:2019wlx}. Taking the $({\cal R}^{\tau}_{\mu})^D$ as an example, we present the allowed parameter regions in Fig.~\ref{fig-4}. One can find that the SM predictions are in good agreement with the current data and the allow regions for NP are rather limited.

\section{Semileptonic Decays} \label{sec:semileptonicdecays}
%=================================
\subsection{Results of SM}
%=================================
Due to the large mass of the $\tau$ lepton, the semileptonic $D$ decays involving $\tau$ lepton are kinematically forbidden. All decays with $e$ and $\mu$ leptons can be divided into four categories, according to the different currents, namely $c\to s e^+\nu_e$, $c\to s \mu^+\nu_\mu$, $c\to d e^+\nu_e$ and $c\to d \mu^+\nu_\mu$. In Table.~\ref{tab:sm-th}, we present the branching fractions of all semileptonic $D$ decays in SM, where the only uncertainties arising from the form factors are considered. The current experimental results \cite{Zyla:2020zbs} are also given for comparison.

\begin{table}[!ht]
\setlength{\abovecaptionskip}{0pt}
\setlength{\belowcaptionskip}{10pt}
\begin{center}
\caption{ Branching fractions for semileptonic $D(D_s)$ decays calculated in the SM using LQCD \cite{Lubicz:2017syv, Lubicz:2018rfs} and LCSR\cite{Wu:2006rd}, and  decay constants of $D(D_s)$ come from LQCD\cite{Aoki:2016frl}. And comparison with the current experimental results given in Ref\cite{Zyla:2020zbs}.}\label{tab:sm-th}
\renewcommand\arraystretch{0.9}
\begin{tabular}{c|lcccc}
\toprule
\toprule
Current& Mode & SM & Experiment\\
\midrule
\multirow{7}{*}{$c\to s e^+\nu_e$}
&$D^0\to K^-e^+\nu_e$ & $(3.49\pm0.23)\times 10^{-2}$ & $(3.542\pm 0.0035)\times 10^{-2}$\\
&$D^+\to \overline{K}^0e^+\nu_e$ & $(8.92\pm0.59)\times 10^{-2}$ &$ (8.73\pm0.10)\times 10^{-2}$\\
&$D^0\to K^{*-}e^+\nu_e$ & $(1.92\pm0.17)\times 10^{-2}$ & $(2.15\pm0.16)\times 10^{-2}$\\
&$D^+\to \overline{K}^{*0}e^+\nu_e$ & $(4.98\pm0.45)\times 10^{-2}$ &$(5.40\pm0.10)\times 10^{-2}$ \\
&$D_s^+\to \phi e^+\nu_e$ & $(2.46\pm0.42)\times 10^{-2}$ &$(2.39\pm0.16)\times 10^{-2}$ \\
&$D_s^+\to \eta e^+\nu_e$ & $(1.55\pm0.33)\times 10^{-2}$ &$(2.29\pm0.19)\times 10^{-2}$ \\
&$D_s^+\to \eta^\prime e^+\nu_e$ & $(5.91\pm1.26)\times 10^{-3}$ &$(7.4\pm1.4)\times 10^{-3}$ \\
\hline
\multirow{7}{*}{$c\to s \mu^+\nu_\mu$}
&$D^0\to K^-\mu^+\nu_\mu$ & $(3.40\pm0.22)\times 10^{-2}$ & $(3.41\pm 0.04)\times 10^{-2}$\\
&$D^+\to \overline{K}^0\mu^+\nu_\mu$ & $(8.69\pm0.57)\times 10^{-2}$ &$ (8.76\pm0.19)\times 10^{-2}$\\
&$D^0\to K^{*-}\mu^+\nu_\mu$ & $(1.81\pm0.16)\times 10^{-2}$ & $(1.89\pm0.24)\times 10^{-2}$\\
&$D^+\to \overline{K}^{*0}\mu^+\nu_\mu$ & $(4.71\pm0.42)\times 10^{-2}$ &$(5.27\pm0.15)\times 10^{-2}$ \\
&$D_s^+\to \phi\mu^+\nu_\mu$ & $(2.33\pm0.40)\times 10^{-2}$ &$(1.90\pm0.50)\times 10^{-2}$ \\
&$D_s^+\to \eta\mu^+\nu_\mu$ & $(1.52\pm0.31)\times 10^{-2}$ &$(2.4\pm0.5)\times 10^{-2}$ \\
&$D_s^+\to \eta^\prime \mu^+\nu_\mu$ & $(5.64\pm1.10)\times 10^{-3}$ &$(11.0\pm5.0)\times 10^{-3}$ \\
\hline
\multirow{9}{*}{$c\to d e^+\nu_e$}
&$D^0\to \pi^-e^+\nu_e$ & $(2.63\pm0.32)\times 10^{-3}$ & $(2.91\pm 0.04)\times 10^{-3}$\\
&$D^+\to \pi^0e^+\nu_e$ & $(3.41\pm0.41)\times 10^{-3}$ &$ (3.72\pm0.17)\times 10^{-3}$\\
&$D^0\to \rho^-e^+\nu_e$ & $(1.74\pm0.25)\times 10^{-3}$ & $(1.77\pm0.16)\times 10^{-3}$\\
&$D^+\to \rho^0 e^+\nu_e$ & $(2.25\pm0.32)\times 10^{-3}$ &$(2.18^{+0.17}_{-0.25})\times 10^{-3}$ \\
&$D^+\to \omega^0e^+\nu_e$ & $(1.91\pm0.27)\times 10^{-3}$ &$(1.69\pm0.11)\times 10^{-3}$ \\
&$D^+\to \eta e^+\nu_e$ & $(0.76\pm0.16)\times 10^{-3}$ &$(1.11\pm0.07)\times 10^{-3}$ \\
&$D^+\to \eta^\prime e^+\nu_e$ & $(1.12\pm0.24)\times 10^{-4}$ &$(2.0\pm0.4)\times 10^{-4}$ \\
&$D_s^+\to K^0 e^+\nu_e$ & $(3.93\pm0.82)\times 10^{-3}$ &$(3.9\pm0.9)\times 10^{-3}$ \\
&$D_s^+\to K^{*0} e^+\nu_e$ & $(2.33\pm0.34)\times 10^{-3}$ &$(1.8\pm0.4)\times 10^{-3}$ \\
\hline
\multirow{9}{*}{$c\to d \mu^+\nu_\mu$}
&$D^0\to \pi^-\mu^+\nu_\mu$ & $(2.60\pm0.31)\times 10^{-3}$ & $(2.67\pm 0.12)\times 10^{-3}$\\
&$D^+\to \pi^0\mu^+\nu_\mu$ & $(3.37\pm0.40)\times 10^{-3}$ &$ (3.50\pm0.15)\times 10^{-3}$\\
&$D^0\to \rho^-\mu^+\nu_\mu$ & $(1.65\pm0.23)\times 10^{-3}$ & $--$\\
&$D^+\to \rho^0 \mu^+\nu_\mu$ & $(2.14\pm0.30)\times 10^{-3}$ &$(2.4\pm0.4)\times 10^{-3}$ \\
&$D^+\to \omega^0 \mu^+\nu_\mu$ & $(1.82\pm0.26)\times 10^{-3}$ &$--$ \\
&$D^+\to \eta\mu^+\nu_\mu$ & $(0.75\pm0.15)\times 10^{-3}$ &$--$ \\
&$D^+\to \eta^\prime\mu^+\nu_\mu$ & $(1.06\pm0.20)\times 10^{-4}$ &$--$ \\
&$D_s^+\to K^0\mu^+\nu_\mu$ & $(3.85\pm0.76)\times 10^{-3}$ &$--$ \\
&$D_s^+\to K^{*0}\mu^+\nu_\mu$ & $(2.23\pm0.32)\times 10^{-3}$ &$--$ \\
\bottomrule
\bottomrule
\end{tabular}
\end{center}
\end{table}

From the table, one can see that all theoretical predictions are in agreement with the current experimental results at $1\sigma$ confidence level. If looking at this table more closely, we find that for the $D\to P \ell^+\nu_\ell$ the differences between the theoretical predictions and the experimental data are very small, except the channels with $\eta$ and $\eta^\prime$. It is known to us that $\eta$ and $\eta^\prime$ are mixing states of $\eta_1$, $\eta_8$ and possible gluonic content. The mixing angels have not been determined yet, which will bring large theoretical uncertainties. The branching fraction of decay $D_s^+\to K^0 \mu^+\nu_\mu$ has not been measured till now, and the order of magnitude is around of the corner in BESIII experiment. For the $D \to V \ell^+\nu_\ell$ decays, the predicted branching fractions of $D^0\to K^{*-} e^+\nu_e$, $D^0\to K^{*-} \mu^+\nu_\mu$, $D^0\to \rho^{-} e^+\nu_e$  and  $D^+\to \rho^{0} e^+\nu_e$ in SM are well in agreement with experimental data, even for center values. However, for the decays $D^+\to \overline{K}^{*0}e^+\nu_e$ and $D^+\to \overline{K}^{*0} \mu^+\nu_\mu$, the SM results are a bit smaller than the data with larger theoretical uncertainties. On the contrary, the predicted branching fractions of $D_s^+\to \phi \mu^+\nu_\mu$ and $D_s^+\to \overline{K}^{*0}e^+\nu_e$ are larger than the experimental data, although there are larger uncertainties in both theoretical and experimental sides. We also note that for the form factors of $D \to K^*$, there are large differences among results calculated within different approaches, which leads to large theoretical uncertainties. For example, for the decays $D^+\to \overline{K}^{*0}e^+\nu_e$ and $D^+\to \overline{K}^{*0} \mu^+\nu_\mu$, the branching fractions based on LQCD \cite{Bowler:1994zr} are lager than those based on LCSR \cite{Wu:2006rd} by $25\%$. Because many form factors of $D \to V$ of LQCD are absent now, we adopt the results of LCSR in order to keep the consistency. In this respect, the reliable calculations of $D \to V$ are needed, especially from Lattice QCD, to match the more precise experimental data.
%=================================
\subsection{Constraints on New Physics}
%=================================
As aforementioned, all decays are induced by $c\to s e^+\nu_e$, $c\to s \mu^+\nu_\mu$, $c\to d e^+\nu_e$ and $c\to d \mu^+\nu_\mu$ currents. In order to study the new physics contributions, we shall discuss two cases. For the case-I, we keep the LFU and assume that the Wilson coefficients of new physics operators are same for muon and electron. While for the case-II, we suppose that NP violates LFU and affects the electron sector and muon sector individually. It should be noted that the differences between $c\to d \ell^+\nu_\ell$ and $c\to s \ell^+\nu_\ell$, namely the violation of light quark $SU(3)$ symmetry, were not discussed any more for simplicity.

Using the existing experimental data including the leptonic and semileptonic decays, and considering the single operator contribution, we can constrain the Wilson coefficients of each new physics operator. In the calculation, we will perform the minimum $\chi^2$ fit of the Wilson coefficients at the $1\sigma$ C.L. of experiment and theory. In our methodology of minimum $\chi^2$ fit, the $\chi^2$ as a function of the Wilson coefficient $C_{X}^\ell$ is defined as \cite{Alok:2017qsi}
\begin{align}
\chi^2(C_{X})=\sum^{data}_{m=1}\frac{[O^{th}_{m}(C_{X}^\ell)-O^{exp}_{m}]^{2}}{\sigma^{2}_{O^{th}_{m}}+\sigma^{2}_{O^{exp}_{m}}},
\end{align}
where $O^{th}_{m}(C_{X}^\ell)$ are the theoretical predictions for different branching fractions, and  $O^{exp}_{m}$ are the corresponding experimental measurements, which are all listed in Table.~\ref{leptonicresult} and \ref{tab:sm-th}. $\sigma_{O^{th}_{m}}$ and $\sigma_{O^{exp}_{m}}$ are the theoretical and experimental errors, respectively. In addition, since there are large theoretical uncertainties in the form factors of $D\to \eta^{(\prime)}$, we will not use the data of $D\to \eta^{(\prime)}\ell \nu_\ell$ in the fitting.

\begin{table}[t]
	\setlength{\abovecaptionskip}{0pt}
	\setlength{\belowcaptionskip}{10pt}
	\begin{center}
		\caption{\label{tab:ExpTh}  Fitted values of the Wilson coefficients for different cases.}\label{fitwcs}
		\renewcommand\arraystretch{1.0}
		\begin{tabular}{c|ccc}
			\toprule
			\toprule
			Case & Wilson Coefficient & Fitted Results &$\chi^{2}_{1\sigma}$ \\
			\midrule
			\multirow{5}{*}{Case-I }
			&$C_{VL}^\ell$ & $(4.3\pm 9.6)\times 10^{-3} $ &$10.1$\\
			&$C_{VR}^\ell$ & $(2.7\pm 9.8)\times 10^{-3}$ &$10.2$\\
			&$C_{SL}^\ell$ & $(0.3\pm 0.6)\times 10^{-3}$ &$10.0$\\
			&$C_{SR}^\ell$ & $(-0.3\pm 0.6)\times 10^{-3}$ &$10.0$\\
			&$C_{T}^\ell$ & $(1.1\pm 2.9)\times 10^{-3} $ &$7.2$\\
			\hline
			\multirow{10}{*}{Case-II}
			&$C_{VL}^e$ & $(9.9\pm 16.2)\times 10^{-3} $ &$4.4$\\
			&$C_{VR}^e$ & $(2.6\pm 16.9)\times 10^{-3} $ &$4.8$\\
			&$C_{SL}^e$ & $(0.4\pm 0.6)\times 10^{-3}$ &$4.8$\\
			&$C_{SR}^e$ & $-(0.4\pm 0.6)\times 10^{-3}$ &$4.8$\\
			&$C_{T}^e$  & $(1.3\pm 3.5)\times 10^{-3} $    &$4.6$\\
			\cline{2-4}
			&$C_{VL}^\mu$ & $(1.4\pm 11.9)\times 10^{-3} $ &$5.5$\\
			&$C_{VR}^\mu$ & $(2.7\pm 12.0)\times 10^{-3} $ &$5.4$\\
			&$C_{SL}^\mu$ & $(70.2\pm 0.6)\times 10^{-3} $ &$3.3$\\
			&$C_{SR}^\mu$ & $-(70.2\pm 0.6)\times 10^{-3}$ &$3.3$\\
			&$C_{T}^\mu$ & $(0.6\pm 4.9)\times 10^{-3}$ &$2.5$\\
			\bottomrule
			\bottomrule
		\end{tabular}
	\end{center}
\end{table}

In Table.~\ref{fitwcs}, we present the fitted Wilson coefficients for two different cases with single operator. For the operators $O_{VL}$ and $O_{VR}$, the related Wilson coefficients $C_{VL}$ and $C_{VR}$ are at the order of ${\cal O}(10^{-3})$ in both two cases. The results of case-II indicate the violation of LFU for the operator $O_{VL}$, but such small effects are buried in the theoretical and experimental uncertainties. For the operators $O_{SL}$ and $O_{SR}$, their Wilson coefficients are also at the order of ${\cal O}(10^{-3})$ in case-I, the effects of which cannot be measured in the current experiments, either. In fact, the most stringent constraints are from the pure leptonic decays. Therefore, in case-II the $C_{SL}^\mu$ and $C_{SR}^\mu$ at the order of ${\cal O}(10^{-2})$ and have rather small uncertainties. However, for the decays $D\to e^+\nu_e$, only the upper limit are available, the fitted $C_{SL}^e$ and $C_{SR}^e$ are at the order of ${\cal O}(10^{-4})$ with very large uncertainties.  As for the Wilson coefficients of the tensor operators that are only constrained by the semi-leptonic $D$ decays, $C_{T}^\ell$ is at the order of ${\cal O}(10^{-3})$ or even small. From above results, all new Wilson coefficients are less than $8\%$, which provides the stringent constraints on new physics models, $W^\prime$ models \cite{He:2017bft,Asadi:2018wea}, leptoquark models \cite{Celis:2012dk, Li:2016vvp, Celis:2016azn}, and models with charged Higgs \cite{Tanaka:1994ay, Iguro:2017ysu, Martinez:2018ynq}. Moreover, we remark that LFU might be violated by the operators $O_{SL}^\ell$ and $O_{SR}^\ell$, which can be further tested in other observables. We acknowledge that our analyses are dependent on the $D\to V$ form factors, and the more precise form factors in future will help us to improve our results.
%=================================
\subsection{Predictions}
%=================================
At first, we shall study the pure leptonic $D$ decays with the electron. As shown in eq.(\ref{pld}), the branching fractions are very sensitive to the operators $O^e_{SL}$ and $O^e_{SR}$, because their contributions are related to $1/m_e$.  With the contribution of $O^e_{SL}$ or $O^e_{SR}$, the branching fractions of these pure leptonic $D$ decays are predicted to be
\begin{eqnarray}\label{rnp}
&&\mathcal{B}(D^+ \rightarrow e^+ \nu_{e}) =(1.6^{+19.6}_{-0.7})\times 10^{-8};\\
&&\mathcal{B}(D_s^+ \rightarrow e^+ \nu_{e}) =(2.4^{+27.6}_{-1.2})\times 10^{-7},
\end{eqnarray}
where the uncertainties are only from the uncertainties of the fitted Wilson coefficients. Comparing the above results with ones of SM in Table.~\ref{leptonicresult}, the current branching fractions are about twice as large as the SM predictions with rather large uncertainties.  According to eq.(\ref{ratio}), one can calculate the ratios as
\begin{eqnarray}\label{}
&&({\cal R}_{\mu}^e)^{D^+}=\frac{\mathcal{B}(D^+ \rightarrow e^+ \nu_{e})}{\mathcal{B}(D^+ \rightarrow \mu^+ \nu_{\mu})_{\rm Ex}}
=(4.3^{+52.4}_{-2.0})\times 10^{-5},\\
&&({\cal R}_{\mu}^e)^{D_s^+}=\frac{\mathcal{B}(D_{s}^+ \rightarrow e^+ \nu_{e})}{\mathcal{B}(D_{s}^+ \rightarrow \mu^+ \nu_{\mu})_{\rm Ex}}
=(4.4^{+50.7}_{-2.1})\times 10^{-5};
\end{eqnarray}
which are larger than the SM predictions
\begin{eqnarray}\label{rsm}
({\cal R}_{\mu}^e)^{D^+}\simeq ({\cal R}_{\mu}^e)^{D_s^+}= 2.3\times 10^{-5}.
\end{eqnarray}
However, the orders of this magnitude are too small to be measured now. We hope the future high intensity experiments can test above results.

\begin{figure*}[!htb]
\begin{center}
\includegraphics[width=7.0cm]{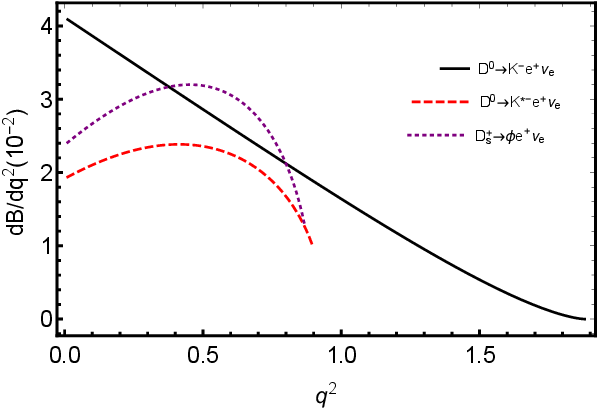}\,\,\,\,\,\,\,\,\,\,\,\,
\includegraphics[width=7.0cm]{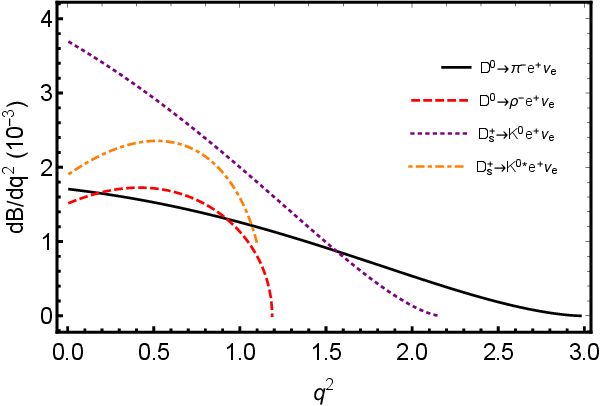}\,\,\,\,\,\,\,\,\,\,\,\,
\includegraphics[width=7.0cm]{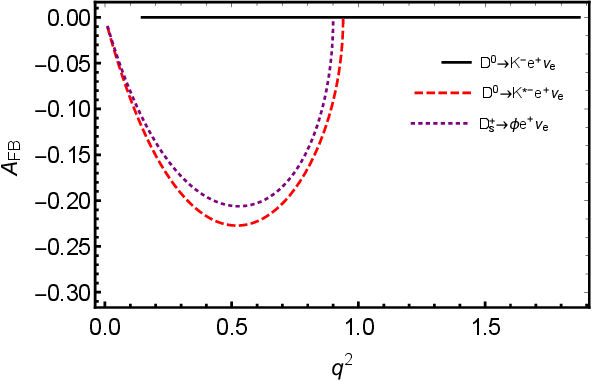}\,\,\,\,\,\,\,\,\,\,\,\,
\includegraphics[width=7.0cm]{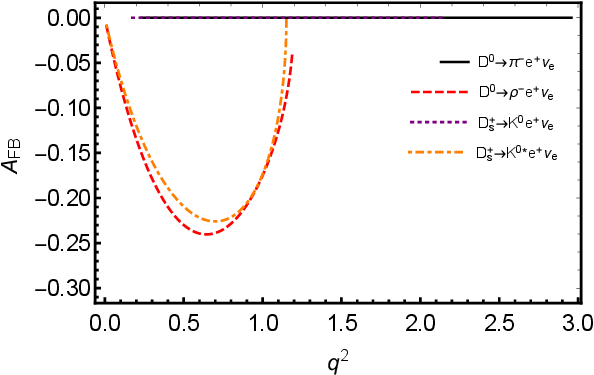}\,\,\,\,\,\,\,\,\,\,\,\,
\includegraphics[width=7.0cm]{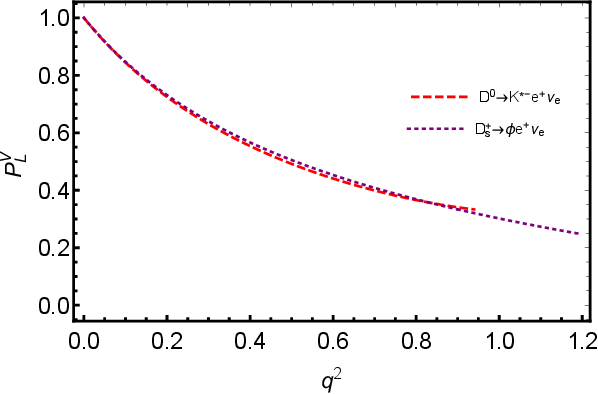}\,\,\,\,\,\,\,\,\,\,\,\,
\includegraphics[width=7.0cm]{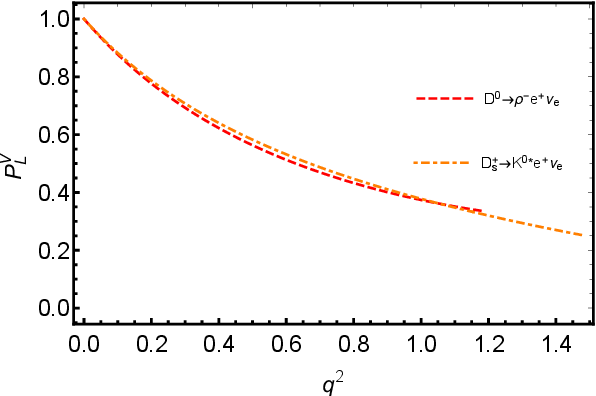}\,\,\,\,\,\,\,\,\,\,\,\,
\includegraphics[width=7.0cm]{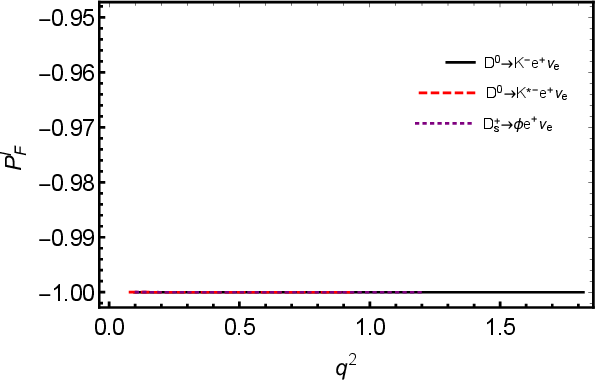}\,\,\,\,\,\,\,\,\,\,\,\,
\includegraphics[width=7.0cm]{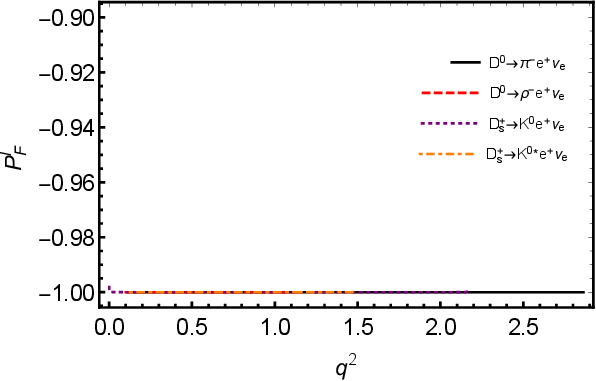}\,\,\,\,\,\,\,\,\,\,\,\,
\caption{The $q^2$-dependence of differential ratios $dBr/dq^2$, the forward-backward asymmetries of the leptonic side $A_{FB}(q^2)$, and the longitudinal polarization components of the vector mesons and electrons in SM.}
\label{fig-5}
\end{center}
\end{figure*}

Secondly, we turn to study the contributions of NP in semileptonic $D$ decays. From Table.~\ref{fitwcs}, it is found that for the decays with electron, the Wilson coefficients in both cases are too small to affect the branching fractions and other observables, such as the differential widths, the forward-backward asymmetries and the polarizations of electrons and final vector mesons. In the experimental sides, all absolute branching fractions have been measured, and the results are in agreement with the SM predictions with uncertainties, as shown in Table.~\ref{tab:sm-th}. However, the $q^2$-dependencies of the differential branching fractions, the forward-backward asymmetries and the polarizations of electrons and final vector mesons have not yet been completely measured now. In Fig.~\ref{fig-5}, we take the decays $D^0 \to K^-e^+\nu_e$,  $D^0 \to K^{*-}e^+\nu_e$, $D_s^+ \to \phi e^+\nu_e$, $D^0 \to \pi^-e^+\nu_e$,  $D^0 \to \rho^-e^+\nu_e$,  $D_s^+ \to K^0e^+\nu_e$ and $D_s^+ \to K^{*0}e^+\nu_e$ as examples and plot the $q^2$-dependencies of the above mentioned observables. Because the mass of electron is negligible, its polarization is almost $-1$. So, any obvious deviations from -1 would be the signals of NP. We hope these predictions can be well tested in BESIII, Belle II and other future high luminosity experiments.

\begin{figure*}[!htb]
\begin{center}
\includegraphics[width=7.0cm]{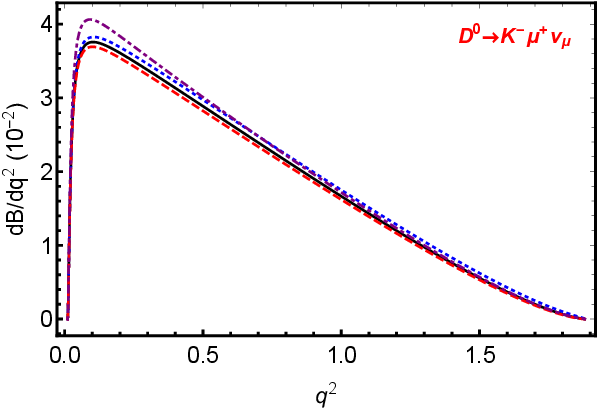}\,\,\,\,\,\,\,\,\,\,\,\,
\includegraphics[width=7.0cm]{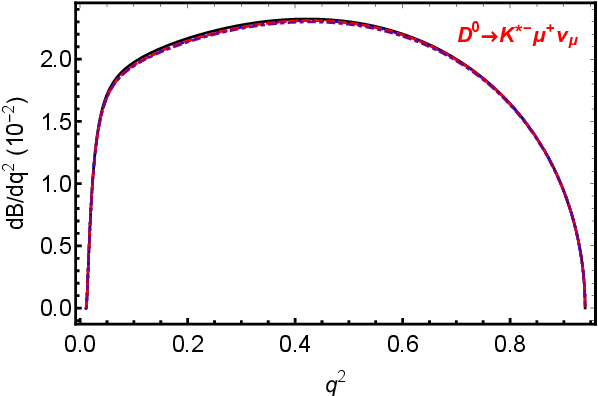}\,\,\,\,\,\,\,\,\,\,\,\,
\includegraphics[width=7.0cm]{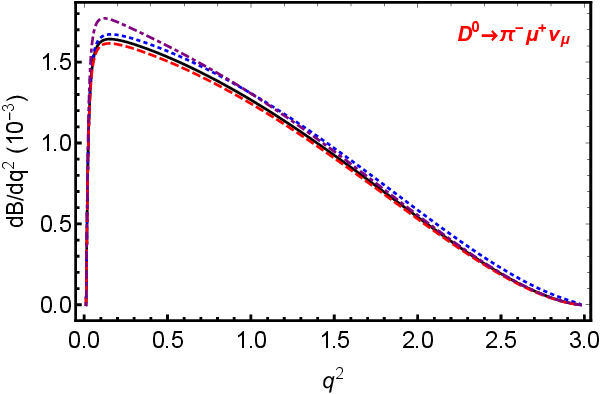}\,\,\,\,\,\,\,\,\,\,\,\,
\includegraphics[width=7.0cm]{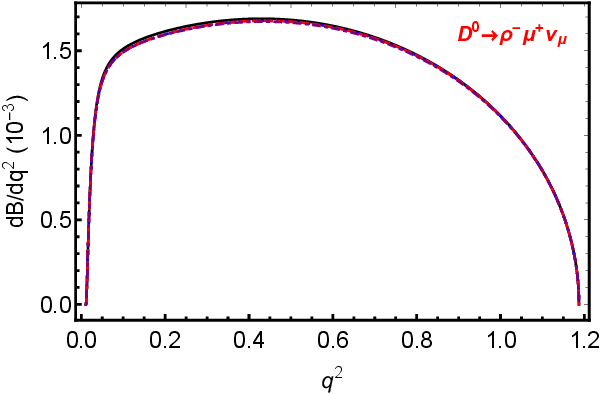}\,\,\,\,\,\,\,\,\,\,\,\,
\includegraphics[width=7.0cm]{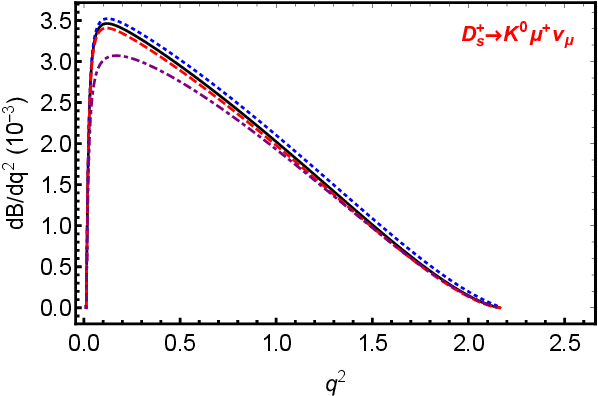}\,\,\,\,\,\,\,\,\,\,\,\,
\includegraphics[width=7.0cm]{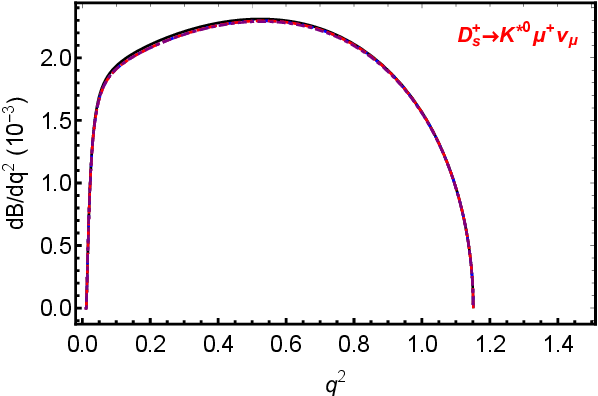}\,\,\,\,\,\,\,\,\,\,\,\,
\includegraphics[width=7.0cm]{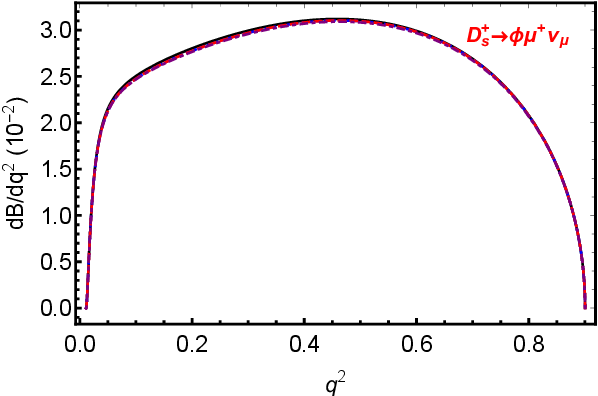}\,\,\,\,\,\,\,\,\,\,\,\,
\caption{The $q^2$-dependence of differential ratios $dBr/dq^2$ of $D^0 \to K^-\mu^+\nu_\mu$, $D^0 \to K^{*-}\mu^+\nu_\mu$, $D^0 \to \pi^-\mu^+\nu_\mu$, $D^0 \to \rho^-\mu^+\nu_\mu$, $D_s^+ \to K^0\mu^+\nu_\mu$, $D_s^+ \to K^{*0}\mu^+\nu_\mu$ and $D_s^+ \to \phi \mu^+\nu_\mu$ with fitted values for decays. The solid (black) lines denote the predictions of SM, while the dotted (blue), dashed (red) and dotdashed (purple) lines  mean NP predictions corresponding to the best-fit Wilson coefficients of ${\cal O}_{SL}$, ${\cal O}_{SR}$, and ${\cal O}_{T}$, respectively.}
\label{fig-6}
\end{center}
\end{figure*}
\begin{figure*}[!htb]
\begin{center}
\includegraphics[width=7.0cm]{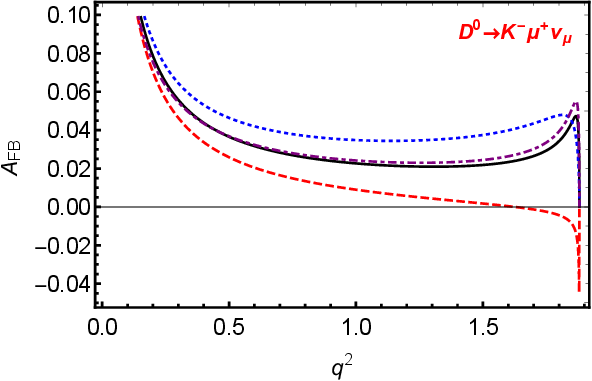}\,\,\,\,\,\,\,\,\,\,\,\,
\includegraphics[width=7.0cm]{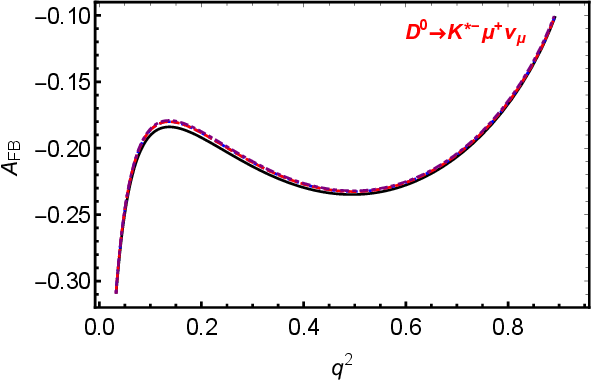}\,\,\,\,\,\,\,\,\,\,\,\,
\includegraphics[width=7.0cm]{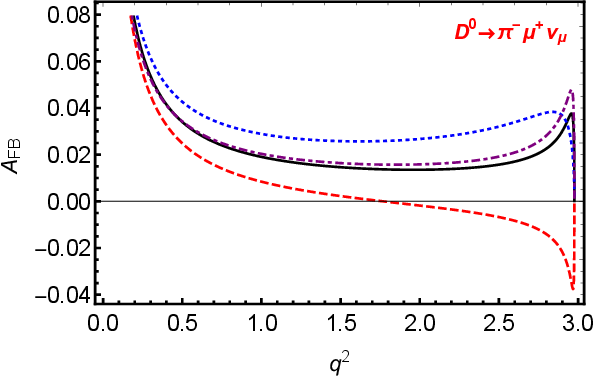}\,\,\,\,\,\,\,\,\,\,\,\,
\includegraphics[width=7.0cm]{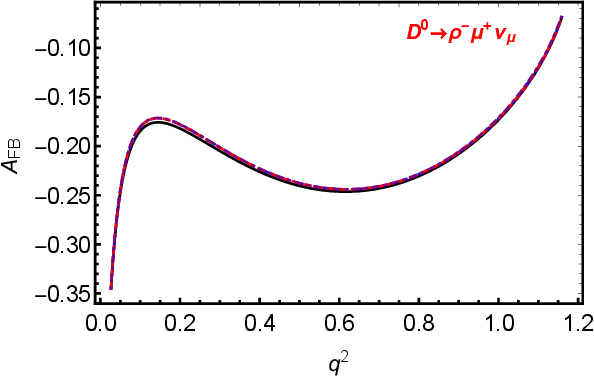}\,\,\,\,\,\,\,\,\,\,\,\,
\includegraphics[width=7.0cm]{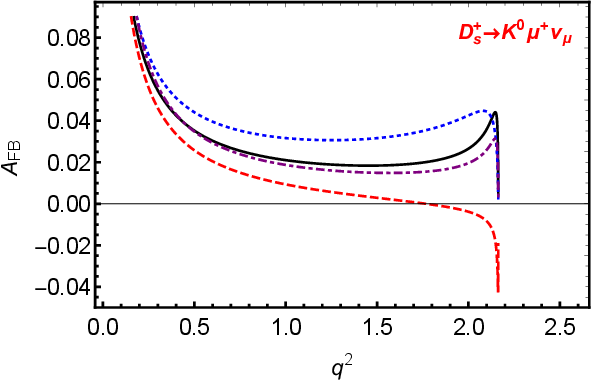}\,\,\,\,\,\,\,\,\,\,\,\,
\includegraphics[width=7.0cm]{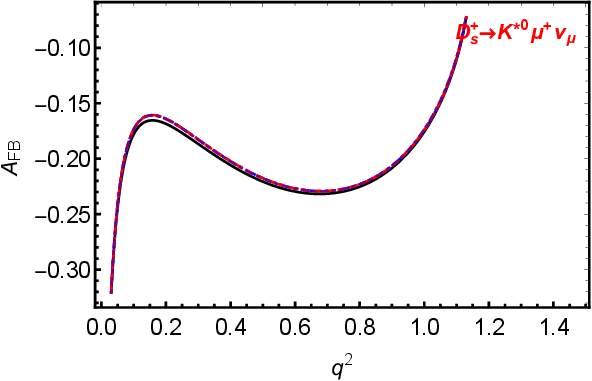}\,\,\,\,\,\,\,\,\,\,\,\,
\includegraphics[width=7.0cm]{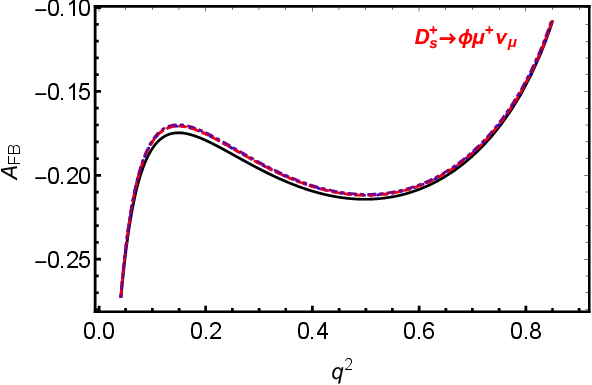}\,\,\,\,\,\,\,\,\,\,\,\,

\caption{The predicted $q^2$-dependence of the forward-backward asymmetries of $D^0 \to K^-\mu^+\nu_\mu$, $D^0 \to K^{*-}\mu^+\nu_\mu$, $D^0 \to \pi^-\mu^+\nu_\mu$, $D^0 \to \rho^-\mu^+\nu_\mu$, $D_s^+ \to K^0\mu^+\nu_\mu$, $D_s^+ \to K^{*0}\mu^+\nu_\mu$ and $D_s^+ \to \phi \mu^+\nu_\mu$ with fitted values for decays. The solid (black) lines denote the predictions of SM, while the dotted (blue), dashed (red) and dotdashed (purple) lines  mean NP predictions corresponding to the best-fit Wilson coefficients of ${\cal O}_{SL}$, ${\cal O}_{SR}$, and ${\cal O}_{T}$, respectively.}
\label{fig-7}
\end{center}
\end{figure*}

\begin{figure*}[!htb]
\begin{center}
\includegraphics[width=7.0cm]{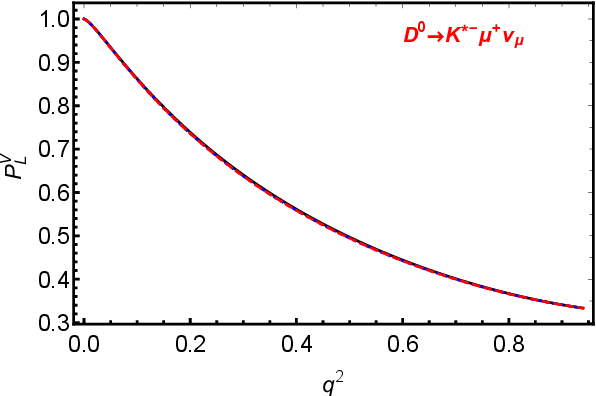}\,\,\,\,\,\,\,\,\,\,\,\,
\includegraphics[width=7.0cm]{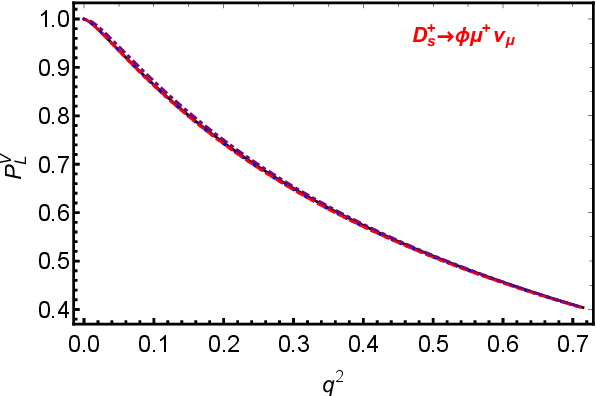}\,\,\,\,\,\,\,\,\,\,\,\,
\includegraphics[width=7.0cm]{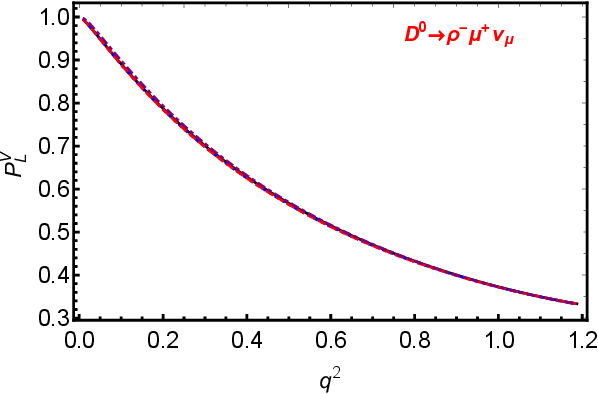}\,\,\,\,\,\,\,\,\,\,\,\,
\includegraphics[width=7.0cm]{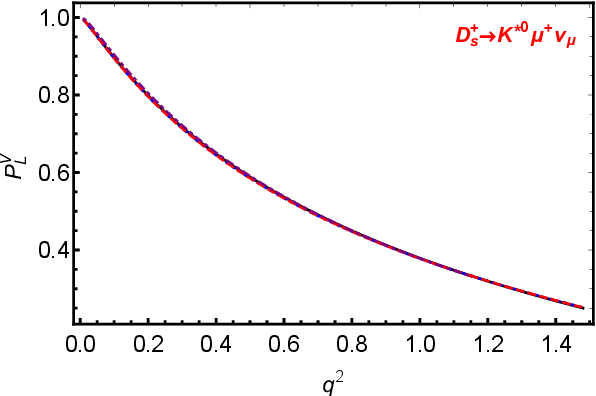}\,\,\,\,\,\,\,\,\,\,\,\,
\caption{The predicted $q^2$-dependence of the polarizations of vector mesons of $D^0 \to K^{*-}\mu^+\nu_\mu$, $D_s^+ \to \phi \mu^+\nu_\mu$, $D^0 \to \rho^-\mu^+\nu_\mu$ and $D_s^+ \to K^{*0}\mu^+\nu_\mu$ with fitted values for decays. The solid (black) lines denote the predictions of SM, while the dotted (blue), dashed (red) and dotdashed (purple) lines  mean NP predictions corresponding to the best-fit Wilson coefficients of ${\cal O}_{SL}$, ${\cal O}_{SR}$, and ${\cal O}_{T}$, respectively.}
\label{fig-8}
\end{center}
\end{figure*}

\begin{figure*}[!htb]
\begin{center}
\includegraphics[width=7.0cm]{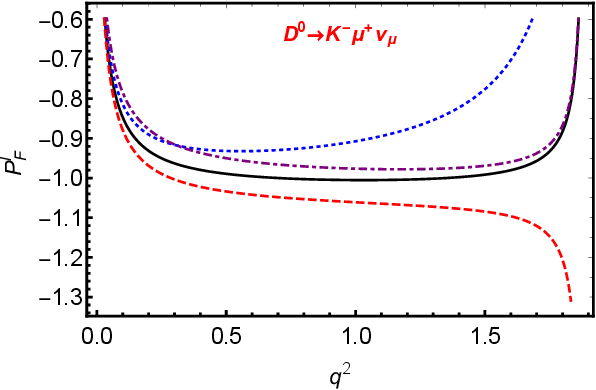}\,\,\,\,\,\,\,\,\,\,\,\,
\includegraphics[width=7.0cm]{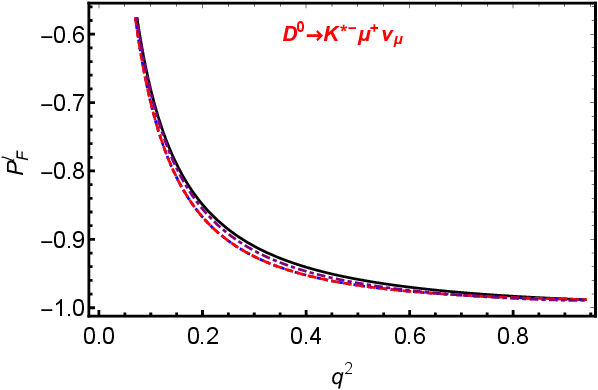}\,\,\,\,\,\,\,\,\,\,\,\,
\includegraphics[width=7.0cm]{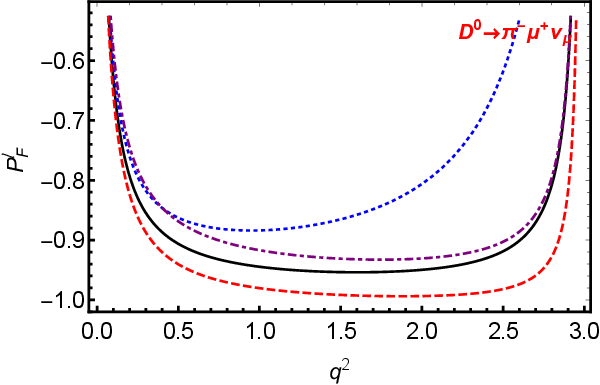}\,\,\,\,\,\,\,\,\,\,\,\,
\includegraphics[width=7.0cm]{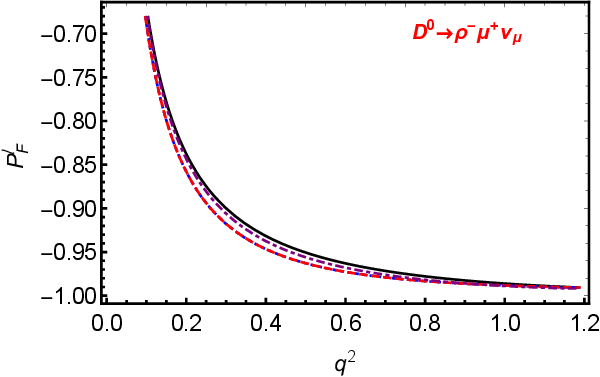}\,\,\,\,\,\,\,\,\,\,\,\,
\includegraphics[width=7.0cm]{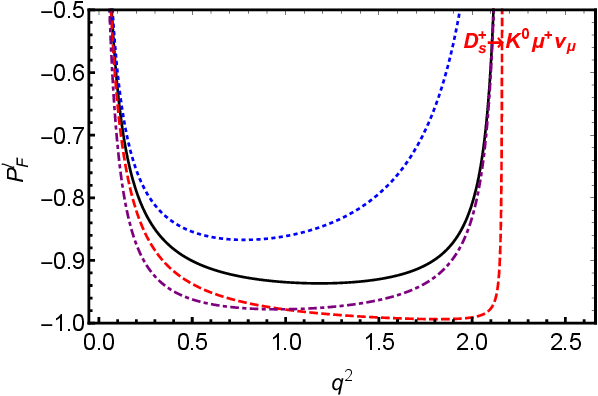}\,\,\,\,\,\,\,\,\,\,\,\,
\includegraphics[width=7.0cm]{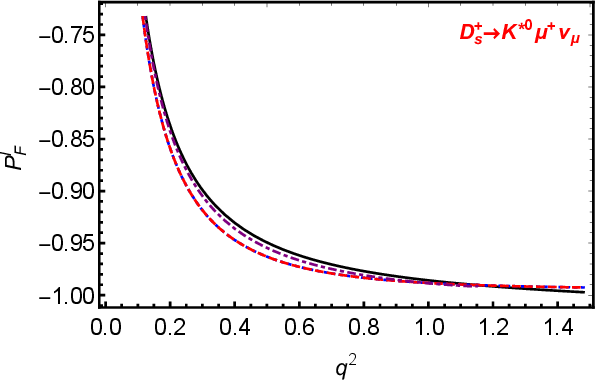}\,\,\,\,\,\,\,\,\,\,\,\,
\includegraphics[width=7.0cm]{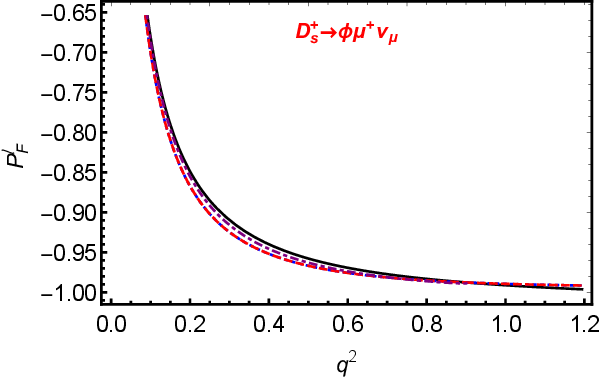}\,\,\,\,\,\,\,\,\,\,\,\,
\caption{The predicted $q^2$-dependence of the lepton hilicity asymmetries of $D^0 \to K^-\mu^+\nu_\mu$, $D^0 \to K^{*-}\mu^+\nu_\mu$, $D^0 \to \pi^-\mu^+\nu_\mu$, $D^0 \to \rho^-\mu^+\nu_\mu$, $D_s^+ \to K^0\mu^+\nu_\mu$, $D_s^+ \to K^{*0}\mu^+\nu_\mu$ and $D_s^+ \to \phi \mu^+\nu_\mu$ with fitted values for decays. The solid (black) lines denote the predictions of SM, while the dotted (blue), dashed (red) and dotdashed (purple) lines  mean NP predictions corresponding to the best-fit Wilson coefficients of ${\cal O}_{SL}$, ${\cal O}_{SR}$, and ${\cal O}_{T}$, respectively.}
\label{fig-9}
\end{center}
\end{figure*}

Lastly, we shall investigate the NP effects in the semileptonic $D$ decays with muon lepton. Again, it was seen from Table.~\ref{fitwcs} that for the processes $c\to (d,s) \mu^+ \nu_\mu$ the Wilson coefficients of the scalar operators ${\cal O}_{SL,SR}$ are at the order of $10^{-2}$, and that of the tensor operator ${\cal O}_T$ can also reach $5.4\times 10^{-3}$. Such small Wilson coefficients cannot change the branching fractions remarkably, but may affects other observables. In order to check their effects, we take the decays $D^0 \to K^-\mu^+\nu_\mu$, $D^0 \to K^{*-}\mu^+\nu_\mu$, $D_s^+ \to \phi \mu^+\nu_\mu$, $D^0 \to \pi^-\mu^+\nu_\mu$, $D^0 \to \rho^- \mu^+ \nu_\mu$, $D_s^+ \to K^0\mu^+\nu_\mu$ and $D_s^+ \to K^{*0}\mu^+\nu_\mu$ as examples and study their contributions to the $q^2$-dependencies of the differential widths, the forward-backward asymmetries and the polarizations of final vector mesons and muon, and the results are shown in Fig.~\ref{fig-6}, Fig.~\ref{fig-7}, Fig.~\ref{fig-8} and Fig.~\ref{fig-9}, respectively. From these figures, one can see that for the decays $D \to V \ell^+\nu_\ell$, the Wilson coefficients of NP hardly affect the observables. Therefore, if large deviations were measured in future, it would be large challenge for us to understand the resource of NP.

For the decays $D \to P \ell^+\nu_\ell$, the differential widths are insensitive to the fitted Wilson coefficients of NP operators, either. On the contrary, the forward-backward asymmetries $A_{FB}(q^2)$ and the polarizations of the muon lepton $P_F^\ell(q^2)$ are very sensitive to the fitted Wilson coefficients, especially to the ones of the scalar operators, as shown in Fig.~\ref{fig-7} and Fig.~\ref{fig-9}. We take the decay $D^0 \to K^-\mu^+\nu_\mu$ induced by $c\to s \mu^+ \nu_\mu$ as an example for illustration. For the forward-backward asymmetry $A_{FB}(q^2)$, it is always positive in SM and the contribution of tensor operator is negligible. In the decay distribution, given in eq.~(\ref{eq:P-differential angular}), $A_3^P$ is dominant in the low $q^2$ region. Because $A_3^P$ is not related to the scalar operators, their contributions are not remarkable. While in the large $q^2$ region, $A_1^P$ becomes important. Because the $A_1^P$ depends on $|C_{SL}+C_{SR}|^2$, the large deviation from SM prediction in the large $q^2$ region is reasonable. Furthermore, it is found that with the fitted $C_{SR}^\mu$ the forward-backward asymmetries of decays $D^0 \to K^-\mu^+\nu_\mu$, $D^0 \to \pi^-\mu^+\nu_\mu$ and $D_s^+ \to K^0\mu^+\nu_\mu$ cross the zero point, when $q^2=1.57~{\rm GeV}^2$, $1.71~{\rm GeV}^2$ and $1.76~{\rm GeV}^2$, respectively. This special behavior can be used to probe the right-handed scalar current. Similarly, for decays  $D^0 \to K^-\mu^+\nu_\mu$, $D^0 \to \pi^-\mu^+\nu_\mu$ and $D_s^+ \to K^0\mu^+\nu_\mu$, when $q^2>0.5~{\rm GeV}^2$, the contributions of scalar operators become important and could affect the polarizations of the muon $P_F^\mu$, as shown in Figure.~\ref{fig-9}. We therefore call our experimental colleagues to measure these observables, so as to search for the possible contributions of NP.

\section{Summary} \label{sec:summary}
Recent anomalies of $B \to D^{(*)}\ell^- \bar\nu_\ell$ imply that NP may appear in the $b\to c \ell^- \bar\nu_\ell$ transitions, it is natural to raise the question about such phenomena in the $D$ decays induced by $c\to (s,d)\ell^+\nu_\ell$ transitions. In the experimental side, current measurements on the charm meson decay observables in which $c\to (s,d)\ell^+\nu_\ell$ transitions occur are consistent with the SM predictions. Such consistency affords us an opportunity to constrain the parameter spaces of NP and to further test NP models beyond SM. In this work, we extended the SM by assuming general effective Hamiltonians describing the $c\to (s,d)\ell^+\nu_\ell$ transitions, which consists of the full set of the four-fermion operators. Within the latest experimental data, we performed a minimum $\chi^2$ fit of the Wilson coefficient corresponding to each operator in two different cases. We found that the Wilson coefficients of scalar operators can be at the order of ${\cal O}(10^{-2})$, and others are at the order of ${\cal O}(10^{-3})$. With the obtained Wilson coefficients, we then provides predictions for the differential branching fractions, the forward-backward asymmetries and polarizations of final vector mesons and leptons with scalar operators. The pure leptonic decays that are very sensitive to the scalar operators can be used to constrain the scalar Wilson coefficients and test models with charged Higgs. For the semileptonic decays with electron, the effects of NP are negligible, and any deviations from SM predictions would be large challenges for SM and its extensions. As for the semileptonic decays with muon, the scalar operators affect the forward-backward asymmetries and polarizations of muon of $D \to P\mu^+ \nu_\mu$. The future measurements on above observables in BESIII and Belle II experiments will help us to test effects of NP and to further test new physics models.
\section*{Acknowledgment}
We thank Dr.Ivan Nisandzic and Prof.Hai-Long Ma for valuable discussions. This work was supported in part by the National Natural Science Foundation of China under the Grants No. 11975195 and 11705159; and the Natural Science Foundation of Shandong province under the Grant No. ZR2019JQ004 and No. ZR2018JL001. This work is also supported by the Project of Shandong Province Higher Educational Science and Technology Program under Grants No. 2019KJJ007.
\bibliographystyle{bibstyle}
\bibliography{mybibfile}
\end{document}